\documentclass[a4paper,oneside,final,notitlepage,onecolumn]{article} 
\usepackage{amsfonts} 
  
\begin{document} 

%%%%%%%%%%%%%%%%%%%%%%%%%%%%%%%%%%%%%%%%%%%%%%%%%%%%%%%%%%%%%%%%%%%%%%%%%%%%%% 
\newtheorem{theorem}{Theorem}[section] 
\newtheorem{lemma}{Lemma}[section] 
\newtheorem{proposition}{Proposition}[section] 
\newtheorem{corollary}{Corollary}[section] 
\newtheorem{conjecture}{Conjecture}[section] 
\newtheorem{example}{Example}[section] 
\newtheorem{definition}{Definition}[section] 
\newtheorem{remark}{Remark}[section] 
\newtheorem{exercise}{Exercise}[section] 
\newtheorem{axiom}{Axiom}[section] 
%%%%%%%%%%%%%%%%%%%%%%%%%%%%%%%%%%%%%%%%%%%%%%%%%%%%%%%%%%%%%%%%%%%%%%%%%%%%%% 
\renewcommand{\theequation}{\thesection.\arabic{equation}} 
% A fenti parancs atdefinialja az egyenleteket szamozo parancsot 
%%%%%%%%%%%%%%%%%%%%%%%%%%%%%%%%%%%%%%%%%%%%%%%%%%%%%%%%%%%%%%%%%%%%%%%%%%%%%% 
  
\bibliographystyle{plain} 
  
\author{Istv\'{a}n R\'{a}cz\thanks{% 
~Fellow of the Japan Society for the Promotion of Science, on leave of absence 
from MTA-KFKI Research Institute for Particle and Nuclear Physics, email: 
istvan@yukawa.kyoto-u.ac.jp} 
\\ %EndAName 
Yukawa Institute for Theoretical Physics\\ 
Kyoto University, Kyoto 606-01, Japan} 
  
\title{On Further Generalization of the Rigidity Theorem for Spacetimes 
with a Stationary Event Horizon or a Compact Cauchy Horizon} 
\maketitle 
  
\begin{abstract} 
A rigidity theorem that applies to smooth electrovac spacetimes which 
represent either (A) an asymptotically flat stationary black hole or (B) a 
cosmological spacetime with a compact Cauchy horizon ruled by closed null 
geodesics was given in a recent work \cite{frw}. Here we enlarge the 
framework of the corresponding investigations by allowing the presence of 
other type of matter fields. In the first part the matter fields are
involved merely implicitly via the assumption that the dominant energy 
condition is satisfied. In the second part Einstein--Klein-Gordon (EKG), 
Einstein--[non-Abelian] Higgs (E[nA]H), 
Einstein--[Maxwell]--Yang-Mills-dilaton (E[M]YMd) and 
Einstein--Yang-Mills--Higgs (EYMH) systems are studied. 
The black hole event horizon or, respectively, the compact Cauchy horizon of 
the considered spacetimes is assumed to be a smooth non-degenerate null 
hypersurface. It is proven that there exists a Killing vector field in a 
one-sided neighborhood of the horizon in EKG, E[nA]H, E[M]YMd and EYMH 
spacetimes. 
This Killing vector field is normal to the horizon, moreover, the associated 
matter fields are also shown to be invariant with respect to 
it. The presented results provide generalizations of the rigidity theorems 
of Hawking (for case A) and of Moncrief and Isenberg (for case B) and, 
in turn, they strengthen the validity of both the black hole rigidity 
scenario and the strong cosmic censor conjecture of classical general relativity. 
\end{abstract} 
  
\section{Introduction} 
\setcounter{equation}{0} 
  
There are two seemingly disconnected areas within general relativity each 
possessing its own `rigidity' theorem. One of these is the determination of 
the possible asymptotic final states of black holes while the other is the 
justification of the strong cosmic censor hypothesis of Penrose \cite 
{penrose} for closed cosmological models. 
The relevant rigidity theorems of Hawking \cite{hawk,HE} and of Isenberg and 
Moncrief \cite{im1,im2} apply to analytic electrovac spacetimes of type A and 
B, respectively. Since the use of the  analyticity assumption is incompatible 
with the concept of causality the generalization of these rigidity results 
from the analytic to the smooth setting is not only of pure mathematical 
interest. The main result of \cite{frw} was that for both type A and B 
electrovac spacetime configurations the existence of a horizon Killing field -- 
alternatively, the rigidity of these spacetimes -- can be proven for the 
case of smooth geometrical setting. Clearly, it is also of obvious interest 
to know whether these results are valid for the case of other 
type of matter fields. The main purpose of this paper is to show that the rigidity 
theorems of \cite{frw} generalize further to various Einstein-matter systems. 
  
\smallskip 
  
Our principal result can be formulated as follows: Consider a smooth 
spacetime which represents either an asymptotically flat stationary black 
hole or a cosmological spacetime with compact Cauchy horizon generated by 
closed null geodesics. Suppose, furthermore, that the black hole 
event or compact Cauchy horizon of these spacetimes is smooth and 
non-degenerate. Then it is shown that in the cases of EKG, E[nA]H,
E[M]YMd and EYMH systems there exists a smooth Killing vector field
defined in a 
one-sided neighborhood of the horizon. This Killing vector field is normal to 
the horizon, moreover, the associated matter fields are also shown to be 
invariant with respect to it. 

\smallskip

In case of Yang-Mills (YM) configurations interesting new phenomena get into play.
Although the spacetime geometry is found to be as regular as in the case of the other 
particular matter fields, in general, there is a `parallelly propagated' blowing up of 
the `adapted' gauge potentials along the generators of the horizon. This irregular behavior, 
however, is balanced entirely in the associated Lie algebra valued 
2-form field and, in turn, in the energy-momentum tensor. In addition, the adapted gauge 
representations are found to belong to either of the following two characteristically 
disjoint classes: There are `preferred gauge representations' which turn to be invariant 
with respect to the `candidate horizon Killing vector field'. The second class is formed by 
non-preferred gauge representations which possess only certain discrete symmetries. Despite 
of their different characters there is a considerable interplay between these two types of 
representations. Whenever a YM filed possesses a non-preferred gauge representation the 
existence of an infinite `crystal' of non-preferred representations can also be shown. 
Each of these crystals of representations is, however, centered by a preferred gauge representation. 

\smallskip 
  
This paper is organized as follows: In the next section, some preliminary notions and 
results are recalled. Sections \ref{sec-cch} and \ref{sec-fgp} are devoted to the investigation 
of the geometrical properties of `elementary spacetime regions' which are covering spaces of 
neighborhoods of sections of the horizons of the considered spacetimes. Section \ref{sec-mf} 
is for the study of EKG, E[nA]H, E[M]YMd and EYMH systems. It is shown there that these 
systems possess a `candidate horizon Killing vector field'. In section \ref{sec-ext}, first 
it is demonstrated that the elementary spacetime regions extend to a spacetime possessing a 
bifurcate null hypersurface. The Lie derivative of the metric and the matter fields with 
respect to the candidate Killing vector field are shown to vanish on this null hypersurface.
Finally, the null characteristic initial value problem is applied to
prove our main result while
section \ref{sec-final} contains our concluding remarks. 
  
\section{Preliminaries} 
  
\setcounter{equation}{0} 
  
In this section the basic notions and results regarding the set of 
spacetimes studied in this paper will be recalled. Throughout this paper, 
unless otherwise stated, a spacetime is considered to be represented by a 
pair $(M,g_{ab})$ where $M$ is a smooth paracompact connected orientable 
manifold while $g_{ab}$ is a smooth Lorentzian metric of signature $(+,-,-,-) 
$ on $M$. It is assumed that $(M,g_{ab})$ is time orientable and also that a 
time orientation has been chosen. 
  
In the first part of this paper the matter fields will merely be involved 
implicitly via the assumption that the dominant energy condition is 
satisfied. Accordingly we shall assume that for all future directed timelike 
vector $\xi ^a$ the contraction ${T^a}_b\xi ^b$ is a future directed 
timelike or null vector, where $T_{ab}$ denotes the energy-momentum tensor. 
  
\smallskip 
  
There will be two classes, $A$ and $B$, of spacetimes investigated 
throughout this paper. In short terms, \textit{class A } (see Sect. 2.1 of 
\cite{frw} for further details) consists of spacetimes which are 
asymptotically stationary with respect to a smooth Killing vector field $t^a$ 
\cite{frw,chw}. The Killing orbits of $t^a$ are assumed to be complete and 
the associated one-parameter group of isometries is denoted by $\phi _t$. The  
event horizon $\mathcal{N}$ is required to be a $\phi_t$-invariant smooth null 
hypersurface so that the manifold of the null geodesic generators of $\mathcal{N}$ 
has the topology $S^2$. Finally, whenever matter fields are present they are 
supposed to be stationary. Correspondingly, whenever the matter fields are 
represented by tensor fields they are assumed to be $\phi_t$-invariant. In case of 
gauge fields, however, only  the existence of an adapted $\phi_t$-invariant gauge 
representation is required. 
  
\smallskip 
  
Spacetimes of \textit{class B} are assumed to possess a compact orientable smooth null 
hypersurface, $\mathcal{N}$, generated by closed null geodesics. In most of the spacetimes
belonging to this class (see e.g. \cite{mill,monc1,monc2}) the null hypersurface $\mathcal{N}$ 
plays the role of a Cauchy horizon. Since the globally hyperbolic region of these spacetimes 
possesses compact Cauchy surfaces spacetimes of class B can be considered, and are frequently 
referred, to be closed cosmological models. 
  
\medskip 
  
It was shown (see Prop. 3.1 of \cite{frw}) that to any spacetime $(M,g_{ab})$ 
of class $A$ there exists an open neighborhood $\mathcal{V}$ of the event horizon  
$\mathcal{N}$ such that $(\mathcal{V},g_{ab})$ is a covering space of a 
spacetime of class $B$. Therefore to give a simultaneous generalization of 
both the results of Hawking and of Isenberg and Moncrief it suffices to show 
the existence of a horizon Killing field for spacetimes of class $B$. 
  
Moreover, for a spacetime of class $B$ the existence of `tubular' spacetime neighborhoods 
$\mathcal{U}_i$ can be shown. These neighborhoods are fibered by circles and sufficiently 
small spacetime neighborhoods of the  Cauchy horizon $\mathcal{N}$ can always be covered 
by a finite subset of them. Most importantly, it 
follows from  the analyses of \cite{frw} that there exist simply connected 
`elementary  spacetime  neighborhoods' $\mathcal{O}_i$ and fiber preserving local  
isometry mappings $\psi_i:\mathcal{O}_i\rightarrow \mathcal{U}_i$ onto the tubular 
spacetime neighborhoods $\mathcal{U}_i$ so that the followings hold: 
  
 $(i)$ Gaussian null coordinates $(u,r,x^3,x^4)$ can be  introduced 
in $\mathcal{O}_i$ such  that the  coordinate range  of $u$  is 
$(-\infty  ,\infty  )$  whereas  the  coordinate  range  of $r$ is 
$(-\epsilon ,\epsilon )$ for some $\epsilon >0$ and the  surface 
$r=0$  is  the  inverse  image  $\widetilde{\mathcal{N}}_i$  of 
$\mathcal{N}_i=\mathcal{N}\cap \mathcal{U}_i$. 
  
 $(ii)$ $k^a= \left(\partial  /\partial u\right)^a$ can be set to be a future    
directed    null    vector    field    normal   to 
$\widetilde{\mathcal{N}}_i$  while  $l^a=\left(\partial  /\partial 
r\right)^a$  is  defined  in  terms  of  the  affine parameter $r$ 
measured along  the null  geodesics starting  orthogonally to  the 
2-dimensional       cross       sections,       $u=const$,        of 
$\widetilde{\mathcal{N}}_i$  with  tangent  $l^a$  satisfying that 
$l^ak_a=1$ throughout $\widetilde{\mathcal{N}} _i$. 
  
$(iii)$ The spacetime metric in $\mathcal{O}_i$ takes the form 
\begin{equation} 
\mathrm{d}s^2=r\cdot f\mathrm{d}u^2+2\mathrm{d}r\mathrm{d}u+2r\cdot h_A 
\mathrm{d}u\mathrm{d}x^A+g_{AB}\mathrm{d}x^A\mathrm{d}x^B  \label{le1} 
\end{equation} 
where $f,h_A$ and $g_{AB}$ are smooth $u$-periodic functions, with a 
period  $P$,  such that $g_{AB}$ is a negative definite $2\times 
2$ matrix. (The 
uppercase Latin indices take everywhere the values $3,4$.) 
  
$(iv)$ In these coordinates the components of the matter field tensors and also that of 
the adapted gauge representations are $u$-periodic `functions' with periodicity length $P$.
\footnote{%
Although, in most of the cases it does not play any role, unless otherwise stated, we 
shall tacitly assume that $P$ is the smallest possible positive period of the considered 
$u$-periodic functions.}
  
\section{The rigidity of the horizon of a spacetime of class B} 
  
\label{sec-cch} \setcounter{equation}{0} 
  
It is known (see e.g. \cite{HE}) that whenever the null convergence 
condition is satisfied the event horizon of a spacetime belonging to class A 
is rigid. By making use of the same energy condition, along with the $u$-periodicity, 
the Cauchy horizon of spacetimes of class B can also shown to be rigid. 
  
\begin{proposition} \label{prop-3.1} 
Let $(M,g_{ab})$ be a spacetime of class B. Then the closed null geodesic generators of 
$\mathcal{N}$ are expansion and shear free, i.e. for any choice of an elementary neighborhood 
$\left( \mathcal{O}_i,g_{ab}\mid _{\mathcal{O}_i}\right)$ we 
have that $\partial g_{AB}/\partial u=0$ throughout $\widetilde{\mathcal{N}}_i$. 
\end{proposition} 
  
\noindent\textbf{Proof}{\ } Let $\left( \mathcal{O}_i,g_{ab}\mid _{\mathcal{O}_i% 
}\right) $ be an elementary neighborhood. Since $k^a=(\partial /\partial u)^a 
$ is null on $\widetilde{\mathcal{N}}_i$ we have 
\begin{equation} 
R_{ab}k^ak^b=8\pi T_{ab}k^ak^b  \label{ein1} 
\end{equation} 
there. In Gaussian null coordinates (\ref{ein1}) reads on  
$\widetilde{\mathcal{N}}_i$ as 
\begin{equation} 
\frac{\partial ^2\left[ \ln \sqrt{g}\right] }{\partial u^2}+\frac{f}
2\frac{\partial \left[ \ln \sqrt{g}\right] }{\partial u}+\frac 
14g^{AG}g^{BD}\left( \frac{\partial g_{AB}}{\partial u}\right) \left( \frac{ 
\partial g_{GD}}{\partial u}\right) +8\pi T_{ab}k^ak^b=0,  \label{ein1g} 
\end{equation} 
where $g:=-\det \left( g_{AB}\right) $ and the $2\times 2$ matrix $g^{AB}$ 
denotes the inverse of $g_{AB}$. Since $g_{AB}$ is negative definite and the 
null energy condition holds (which follows from the dominant energy 
condition), both of the last two terms on the l.h.s. of (\ref{ein1g}) have 
to be greater than or equal to zero. Moreover, since the metric functions 
$g_{AB}$ are $u$-periodic there exists a point $u_0$ with $(\partial [\ln \sqrt{g}
]/\partial u)(u_0)=0$. Reading (\ref{ein1g}) as an equation of first order 
for $\partial [\ln \sqrt{g}]/\partial u$ along the generators of $\widetilde{ 
\mathcal{N}}_i$ we get 
\begin{equation} 
\left( \frac{\partial [\ln \sqrt{g}]}{\partial u}\right) (u)=-e^{-\frac 
12\int_{u_0}^u{f(u^{*})\,du^{*}}}\int_{u_0}^ub(u^{\prime })e^{\frac 
12\int_{u_0}^{u^{\prime }}{f(u^{\prime \prime })}\,du^{\prime 
\prime }}\,du^{\prime },  \label{2gAB} 
\end{equation} 
where $b$ stands for the last two terms of the l.h.s. of (\ref{ein1g}). 
Using again the periodicity, along with the fact that $b\geq 0$, we find 
that both $\partial \left[ \ln \sqrt{g}\right] /\partial u$ and $b$ have to 
vanish identically along the generators of $\widetilde{\mathcal{N}}_i$. \footnote{% 
I wish to thank Helmut Friedrich for this elementary argument demonstrating the vanishing 
of the functions $\partial \left[\ln \sqrt{g}\right] /\partial u$ and $b$.} Thereby, both 
of the last two terms on the l.h.s. of (\ref{ein1g}) have also to vanish identically on 
$\widetilde{\mathcal{N}}_i$. Hence we obtain that\footnote{% 
As a shorthand way of notation we shall denote by $\varphi ^{\circ }$ the restriction 
$\varphi\mid_{\widetilde{\mathcal{N}}_i}$ of a function $\varphi$ to $\widetilde{\mathcal{N}}_i$.} 
\begin{equation} 
\left( \frac{\partial g_{AB}}{\partial u}\right) ^{\circ }=0 \label{mn}
\end{equation} 
holds. \hfill \fbox{} 
  
\begin{remark}\label{rem-gs}
It follows from (\ref{mn}), along with the dominant energy condition, that $k^a$ is a repeated 
principal null vector of the Weyl tensor, i.e. $k_{[a}C_{b]cdf}k^ck^f\equiv 0$ on 
$\widetilde{\mathcal{N}}_i$. To see this note first that in virtue of (\ref{mn}) the spin 
coefficients $\lambda$ and $\mu$  must vanish on 
$\widetilde{\mathcal{N}}_i$. Moreover, the dominant energy condition implies that $R_{ab}k^aX^b=0$ 
(see e.g. the first part of the proof of Prop. \ref{prop-4.1}) for any vector field $X^b$ tangent 
to $\widetilde{\mathcal{N}}_i$. Hence, the Ricci spinor components $\Phi _{22}$ and $\Phi _{21}$ 
also have to be zero throughout $\widetilde{\mathcal{N}}_i$. Finally, by making use of (NP.10), 
(NP.13) and (NP.14) [see footnote 6] the vanishing of $\Psi_3^\circ$ and $\Psi_4^\circ$ can be justified.
\end{remark} 
  
\begin{remark} 
Contrary to the general expectations (see e.g. section 8.5 of \cite{HE}) whenever the dominant 
energy condition is satisfied  $k_{[a}R_{b]cdf}k^ck^f$ vanish identically along the closed null 
geodesic generators of a compact Cauchy horizon. Nevertheless, some (other) components of the 
curvature tensor, in parallelly propagated frames, can blow up there (see e.g. Remark \ref{rem-gs2}). 
\end{remark} 
  
\begin{remark} 
There is another possible reading of the above result:  In a spacetime  satisfying both  the 
`genericness  condition' and the dominant  energy  condition  there  cannot  exist a smooth 
compact Cauchy horizon ruled by closed null geodesics. 
\end{remark} 
  
\section{Further properties of elementary neighborhoods}\label{sec-fgp} \setcounter{equation}{0} 
  
 By  making  use  the  dominant  energy  condition, along with the 
$u$-periodicity, the following can be proven: 
  
\begin{proposition} 
\label{prop-4.1} Let $(\mathcal{O}_i,g_{ab}\vert_{\mathcal{O}_i})$ be an 
elementary neighborhood of $\widetilde{\mathcal{N}}_i$ such that 
$(\partial 
g_{AB}/\partial u)^\circ$ is identically zero. Then there always exists a 
Gaussian null coordinate system so that $f^{\circ}=-2\kappa _{\circ }$, with 
a constant $\kappa_{\circ }\ge 0$, and that $\partial h_A/\partial u=0$ 
throughout $\widetilde{\mathcal{N}}_i$. 
\end{proposition} 
  
\noindent\textbf{Proof}{\  }  Since  $k^a  $  is  normal  to  the 
coordinate basis field $\left(\partial /\partial x^A\right)^a$  on 
$\widetilde{\mathcal{N}}_i$ 
\begin{equation} 
R_{ab}k^a\left(\frac{\partial }{\partial x^A}\right)^b= 8\pi T_{ab}k^a\left(% 
\frac{\partial }{\partial x^A}\right)^b  \label{ein2} 
\end{equation} 
there. Moreover, in virtue of the dominant energy condition 
${T^a}_bk^b$ 
has to be a future directed timelike or null vector. 
On the other hand, 
$T_{ab}k^ak^b=0$ on $\widetilde{\mathcal{N}}_i$. This implies that 
 ${T^a}_{b}k^b$ must point in the direction of $k^a,$ i.e. 
$T_{ab}k^a\left(\partial 
/\partial x^A\right)^b=0$ on $\widetilde{\mathcal{N}}_i$. Hence, by making use 
of the fact that $\left(\partial g_{AB}/\partial u\right)^{\circ}=0$, we get 
that in the underlying Gaussian null coordinates (\ref{ein2}) reads as 
\begin{equation} 
\left( \frac{\partial f}{\partial x^A}-\frac{\partial h_A}{\partial u} 
\right) ^{\circ}=0.  \label{ein2g} 
\end{equation} 
Integrating this equation with respect to $u$ and using the $u$-periodicity 
and smoothness of $h_A$ we get that 
\begin{equation} 
\frac \partial {\partial x^A}\int_0^{P}f^{\circ }\left(u,x^3,x^4\right) 
\mathrm{d}u=0, 
\end{equation} 
which in particular means that for some constant $\kappa_{\circ }\geq 0$ 
\begin{equation} 
\int_0^{P}f^{\circ }(u,x^3,x^4)\mathrm{d}u=-2P \kappa _{\circ } 
\label{kappa} 
\end{equation} 
holds throughout $\widetilde{\mathcal{N}}_i$. (If $\kappa _{\circ }$ was 
smaller than zero then we could achieve $\kappa _{\circ }\geq 0$ by 
the application of the transformation $(u,r,x^3,x^4)\rightarrow 
(-u,-r,x^3,x^4)$ along with the simultaneous reversing of the 
time orientation.) 
  
To prove then that new Gaussian null coordinates can be introduced so that 
$f^{\circ}\equiv-2\kappa_{\circ }$ holds the argument of Moncrief and Isenberg 
(see pages 395-398 of \cite{im1}) can be applied. Since in these new coordinates 
$f^{\circ }$ is constant the relevant form of 
(\ref{ein2g}) implies that $(\partial h_A/\partial u)^{\circ }=0.$ \hfill \fbox{}

\begin{remark}  According  to  (\ref{kappa})  in  any  elementary 
neighborhood the $u$-periodicity of the metric functions  selects 
a  preferred  value  of  $\kappa  _{\circ  }$  which  is  uniquely 
determined in  any associated  Gaussian null  coordinate system as 
$\kappa _{\circ }=-{1/2P}\int^{P}_0 f^{\circ }du$.  Then,  because 
$k^a=(\partial/\partial   u)^a$  on   $\widetilde{\mathcal{N}}_i$ 
satisfies   
\begin{equation}   k^a\nabla_ak^b=\kappa_{\circ   }k^b, \label{sg}  
\end{equation}  
we  have  that  the null generators of 
$\widetilde{\mathcal{N}}_i$ --  which are  complete with  respect to 
the parameter  $u$ --  are null  geodesically complete  if $\kappa 
_{\circ }=0,$ whereas, if $\kappa _{\circ }$ happens to be nonzero 
the  generators  of  $\widetilde{\mathcal{N}}_i$  are   geodesically 
complete only in one direction. \end{remark} 
  
 \begin{remark}\label{rem-zero}
 There  is a  significant consequence  of Prop.  \ref{prop-4.1}
in  connection  with  the  `zeroth  law'  of  black   hole 
thermodynamics.  It is known that for any static  black 
hole  (in  an  arbitrary  covariant  metric theory of gravity) the 
surface gravity has to  be constant throughout the  event horizon \cite{rw2}. 
In view of the above result there is a unique way to introduce a 
quantity,  $\kappa_{\circ}$, which plays the role of surface gravity. 
The constancy of $\kappa_{\circ}$ throughout $\mathcal{N}$ is 
guaranteed \footnote{% 
Actually, the above argument  is `local' in the  space directions 
but, in virtue of (\ref{kappa}),  it is obvious that the  value of 
$\kappa  _{\circ  }$  has  to  be  the same on overlapping primary 
neighborhoods.} by the dominant energy condition.
\end{remark} 
 
\begin{remark}
Based on Prop. \ref{prop-3.1}  and  Prop. \ref   {prop-4.1} hereafter, 
without loss of generality, we shall assume that the 
Gaussian null  coordinate system  $(u,r,x^3,x^4)$ associated  with 
an  elementary  spacetime  region $\mathcal{O}_i$ is so that $f^\circ=-2\kappa 
_{\circ   }$, moreover, $ h_A$ and $g_{AB}$ are $u$-independent on 
$\widetilde{\mathcal{N}}_i$. 
\end{remark}

\smallskip 
 
 For particular spacetimes further  characterization of  the functions 
$f,h_A$ and  $g_{AB}$ can  also be  given.  For  instance, for the 
case of an asymptotically flat stationary non-static vacuum  black 
hole it was  proven in \cite{HE}  that the $r$-derivatives  of the 
functions   $f,   h_A$   and   $g_{AB}$   up   to   any  order are 
$u$-independent on $\widetilde{\mathcal{N}}_i$.  Exactly the  same 
property of these functions was proven to be held for the case  of 
electrovac  cosmological   situations  \cite{im1,im2},   i.e.  for 
spacetimes  possessing  a  compact  Cauchy  horizon  with   closed 
generators  and  satisfying  the  Einstein-Maxwell  equations.  In 
fact,  the  $u$-independentness  of  the  $r$-derivatives  of  the 
functions  $f,h_A$  and  $g_{AB}$  up  to  any  order  is the very 
property  that  can  be  used  to  argue,  in the case of analytic 
spacetime configurations, that $k^a=(\partial/\partial u)^a$ is  a 
Killing   vector    field   in    an   elementary    neighborhood 
${\mathcal{O}_i}$ with respect to which $\widetilde{\mathcal{N}}_i$ is 
a Killing horizon. 
  
\medskip 
  
 The purpose of  the remaining part of this section is to investigate what are  
the  necessary  and   sufficient  conditions  ensuring   that  the 
$r$-derivatives of the functions $f,  h_A$ and $g_{AB}$ up to  any 
order  are  $u$-independent  on $\widetilde{\mathcal{N}}_i$.  Recall 
that the  proof of  Moncrief and  Isenberg \cite{im1,im2}  for the 
electrovac  case  was  based  on  the  detailed application of the 
coupled Einstein-Maxwell  equations.  To  be able  to separate all 
the conditions  which are  in certain  sense `purely  geometrical' 
from the ones which are related to particular properties of matter 
fields  first  we  consider the relevant necessary and 
sufficient  geometrical  conditions.   The  corresponding analysis 
when  the  effect  of  particular  matter  fields  are  taken into 
consideration will be presented in the next section. 
  
 To  prove  the  main  result  of  this  section  the  use  of the 
Newman-Penrose  formalism   \cite{np}  turned   out  to   be  most 
effective.  Thereby, first we recall the relation between the  two 
geometrical settings  based on  the Gaussian  null coordinates and 
the Newman-Penrose formalism, respectively. 
  
 The contravariant form  of the metric  (\ref{le1}) in a  Gaussian 
null coordinate  system $(u,r,x^3,x^4)$,  covering the  elementary 
region ${\mathcal{O}_i}$, reads as 
\begin{equation} 
g^{\alpha\beta}=\left( 
\begin{array}{ccc} 
0 & 1 & 0 \\ 
1 & g^{rr} & g^{rB} \\ 
0 & g^{Ar} & g^{AB} 
\end{array} 
\right) .  \label{m2} 
\end{equation} 
Choosing now real-valued functions $U $, $X ^A$ and complex-valued 
functions $\omega ,$ $\xi ^A$ on ${\mathcal{O}_i}$ such that 
\begin{equation} 
g^{rr}=2(U -\omega \bar\omega),\ \ g^{rA}=X ^A- (\bar\omega\xi 
^A+\omega \bar\xi^A),\ \ g^{AB}=-(\xi ^A \bar\xi^B+\bar\xi^A\xi ^B), 
\label{m3} 
\end{equation} 
and setting 
\begin{equation} 
l^\mu =\delta ^\mu {}_r,\ \ n^\mu =\delta ^\mu {}_u+U \delta ^\mu 
{}_r+X ^A\delta ^\mu {}_A, \ \ m^\mu =\omega \delta ^\mu {}_r+\xi ^A\delta 
^\mu {}_A,  \label{tet} 
\end{equation} 
we obtain a complex null tetrad $\{l^a,n^a,m^a,\overline{m}^a\}$. We require 
that $U $, $X ^A$, and $\omega $ vanish on $\widetilde{\mathcal{N}}_i$ 
such that $n^a$, $m^a$ and $\overline{m}^a$ are tangent to $\widetilde{
\mathcal{N}}_i$. In the following we shall consider the derivatives of 
functions in the direction of the frame vectors above and denote the 
corresponding operators in $\mathcal{O}_i$ by 
\begin{equation} 
\mathrm{D}=\partial /\partial r,\ \ \Delta =\partial /\partial u+ U \cdot
\partial /\partial r+X^A \cdot\partial /\partial x^A, \delta 
=\omega \cdot\partial /\partial r+\xi ^A\cdot\partial /\partial x^A. 
\label{dop} 
\end{equation} 
To simplify the Newman-Penrose equation we fix the remaining gage freedom by 
assuming that the tetrad $\{l^a,n^a,m^a,\overline{m}^a\}$ is parallelly 
propagated along the null geodesics with tangent $l^a=\left( \partial 
/\partial r\right) ^a$. This condition ensures e.g. that for the 
spin coefficients related to this 
complex null tetrad $\kappa =\pi =\varepsilon =0$, $\rho = \overline{\rho }$,
$\tau =\overline{\alpha }+\beta $ hold in $\mathcal{O}_i$, and in 
particular, $\nu =0$, $\gamma =\overline\gamma$ and $\mu =\overline{\mu }$ 
on $\widetilde{\mathcal{N}}_i$.
  
\begin{proposition} 
\label{prop-npind} Denote by $g^{\alpha\beta}$ the contravariant components 
of the spacetime metric in a Gaussian null coordinate system associated with 
an elementary spacetime region $(\mathcal{O}_i,g_{ab}\vert_{\mathcal{O}_i})$. Then 
we have that for all values of $i\in \{1,2,...,n\}$ 
\begin{equation} 
\Delta\left(\mathrm{D}^{(i)}\left(\{g^{\alpha\beta}\}\right)\right)^\circ=0 
\label{dD} 
\end{equation} 
if and only if \footnote{% 
$\Delta\left(\mathrm{D}^{(j)}\left(\{f_1,f_2,...,f_N\}\right)\right)$ denotes the list of 
functions resulted by the action of the differential operators D $(j)$-times and $\Delta$ once 
on the functions $f_1,f_2,...,f_N$. In particular, $\Delta\left(\mathrm{D}^{(0)}\left(
\{f_1,f_2,...,f_N\}\right)\right)$ is defined to be $\Delta\left(\{f_1,f_2,...,f_N\}\right)$.} 
$\Delta\left(\{(\Phi_{11}+3\Lambda),\Phi_{02}\}\right)^\circ=0 
$ and for all values of $j\in \{0,1,2,...,n-2\}$ 
\begin{equation} 
\Delta\left(\mathrm{D}^{(j)}\left(\{\Phi_{00},\Phi_{01},(\Phi_{11}-3\Lambda) 
,\mathrm{D}\left(\Phi_{02}\right)\}\right)\right)^\circ= 0.  \label{dD2} 
\end{equation} 
\end{proposition} 
  
\noindent\textbf{Proof}{\ } By (\ref{m2}) and (\ref{m3}), along with the vanishing of the 
functions $\omega, X^A, U$ on $\widetilde{\mathcal{N}}_i$, we have that (\ref{dD}) 
is satisfied whenever for all values of $i\in \{1,2,...,n\}$ 
\begin{equation} 
\Delta\left(\mathrm{D}^{(i)}\left(\{\xi^A,\omega,X^A,U\}\right)\right) ^\circ=0.  
\label{dD3} 
\end{equation} 
Moreover, by making use of the metric equations -- see eqs. (6.10a) -
(6.10h) of \cite{np} --  
it can be checked that (\ref{dD3}) holds if and only if for 
all values of $i\in \{0,1,2,...,n-1\}$ 
\begin{equation} 
\Delta\left(\mathrm{D}^{(i)}\left(\{\rho,\sigma,\tau, 
(\gamma+\bar\gamma)\}\right)\right)^\circ=0.  \label{dD4} 
\end{equation} 
Furthermore, since $T_{uu}^\circ=T_{uA}^\circ={(\partial 
g_{AB}/\partial u)}^\circ=0$ we have that $\Phi_{22}^\circ=\Phi_{21}^\circ=
\lambda^\circ=\mu^\circ=0$ as well as that $\Psi_{3}^\circ=\Psi_{4}^\circ=0$ 
(for the later relations see e.g. the argument applied at Remark \ref{rem-gs}). 
  
To see that for $n=1$ (\ref{dD4}) is equivalent to the $u$-independentness 
of $\Phi_{11}+3\Lambda$ and $\Phi_{02}$ on $\widetilde{\mathcal{N}}_i$ note 
first that by the definition of $\gamma$ and (\ref{sg}) 
\begin{equation} 
(\gamma+\bar\gamma)^\circ=(n^an^b\nabla_al_b)^\circ=(k^ak^b\nabla_al_b)^% 
\circ= -(k^al^b\nabla_ak_b)^\circ=-\kappa_\circ, 
\end{equation} 
where $\kappa_\circ=const$ throughout $\widetilde{\mathcal{N}}_i$. Thereby, 
along with the fact that $\bar\gamma^\circ=\gamma^\circ$, we get 
\begin{equation} 
\delta(\gamma)^\circ=\bar\delta(\gamma)^\circ=\Delta(\gamma+\bar\gamma)^% 
\circ =0.  \label{Dg} 
\end{equation} 
To see that $\Delta(\tau)^\circ=0$ note that $\tau=\bar\alpha+\beta$ and by 
(NP.15) \footnote{% 
Throughout this proof the equations referred as `$(NP.n)$' and 
`$(B.m)$' are yielded by the substitution of the above gauge
fixing relations into the `$n^{th}$' Newman-Penrose equation and the
`$m^{th}$' Bianchi identity of the Newman-Penrose formalism as they are
listed in the appendix of \cite{stew}.} 
, (NP.18) and (\ref{Dg}) we have 
\begin{equation} 
\Delta(\alpha)^\circ=\Delta(\beta)^\circ=0.  \label{Dab} 
\end{equation} 
An immediate consequence of (\ref{Dab}) and (NP.12) is that 
\begin{equation} 
\Delta(\Psi_2+2\Lambda)^\circ-\Delta(\Phi_{11}+3\Lambda)^\circ=0. 
\label{Df2} 
\end{equation} 
In virtue of (\ref{Df2}) and (NP.17) $\Delta(\rho)$ satisfy 
\begin{equation} 
\Delta\left(\Delta\left(\rho\right)\right)+\kappa_\circ\Delta(\rho)=0 
\label{Dr2} 
\end{equation} 
on $\widetilde{\mathcal{N}}_i$, with the only periodical solution $% 
\Delta(\rho)^\circ=0$, if and only if $\Delta(\Phi_{11}+3\Lambda)^\circ=0$. 
  
Finally, by making use of (NP.16) it can be shown that $\Delta(\sigma)^\circ=0$ whenever 
$\Delta(\Phi_{02})^\circ=0$. 
  
\medskip 
  
To see that our statement is true for $n=2$ one can proceed as follows. By 
(NP.6) we have that $\Delta\left(\mathrm{D}\left(\gamma+\bar\gamma\right)% 
\right)^\circ=0$ if and only if $\Delta(\Psi_2+\bar\Psi_2-2\Lambda+2% 
\Phi_{11})^\circ=0$. However, in virtue of (\ref{Df2}) the last equation is 
equivalent to $\Delta(\Phi_{11})^\circ=0$ which follows from our assumption 
that both $\Delta(\Phi_{11}+3\Lambda)$ and $\Delta(\Phi_{11}-3\Lambda)$ 
vanish on $\widetilde{\mathcal{N}}_i$. 
  
The vanishing of $\Delta\left(\mathrm{D}\left(\alpha\right)\right)^\circ$, $% 
\Delta\left(\mathrm{D}\left(\beta\right)\right)^\circ$ and, thereby also, of 
$\Delta\left(\mathrm{D}\left(\tau\right)\right)^\circ$ by (NP.3) - (NP.5) and 
our assumptions are equivalent to the condition $\Delta(\Psi_{1})^\circ=0$. 
However, by (B.4) we have that on $\widetilde{\mathcal{N}}_i$ 
\begin{equation} 
\Delta\left(\Delta\left(\Psi_{1}\right)\right)+\kappa_\circ\Delta(\Psi_{1}) 
-\tau\Delta\left(2\Phi_{11}-3\Psi_2\right)=0.  \label{DDs1} 
\end{equation} 
The last term on the l.h.s. of (\ref{DDs1}) is just $\tau\Delta\left(% 
\Phi_{11}+3\Lambda\right)$ which is identically zero on $\widetilde{\mathcal{% 
N}}_i$. Therefore, the only periodical solution of (\ref{DDs1}) is $% 
\Delta(\Psi_{1})^\circ=0$. 
  
It follows immediately from (NP.1) that $\Delta\left(\mathrm{D}% 
\left(\rho\right)\right)^\circ=0$ if and only if $\Delta(\Phi_{00})^\circ=0$% 
, and similarly, from (NP.2) that $\Delta\left(\mathrm{D}\left(\sigma\right)% 
\right)^\circ=0$ precisely when $\Delta(\Psi_{0})^\circ=0$. By our 
assumption we have that $\Delta(\Phi_{00})^\circ=0$ while from (B.2) we get 
that on $\widetilde{\mathcal{N}}_i$ 
\begin{equation} 
\Delta\left(\Delta\left(\Psi_{0}\right)\right)+2\kappa_\circ\Delta(\Psi_{0}) 
-\sigma\Delta\left(3\Psi_2+2\Phi_{11}\right) +\Delta\left(\mathrm{D}% 
\left(\Phi_{02}\right)\right)=0.  \label{DDs2} 
\end{equation} 
Again it follows from our assumptions that the last two terms on the 
l.h.s. of (\ref{DDs2}) vanish, whence the only $u$-periodic solution of (\ref 
{DDs2}) satisfies $\Delta(\Psi_{0})^\circ=0$. 
  
An immediate further consequence of the $u$-independentness of $% 
\alpha,\beta,\gamma,\tau,\rho,\sigma$ and $\Phi_{00},\Phi_{01},\Phi_{11}$ on 
$\widetilde{\mathcal{N}}_i$ are the following: By (B.9) - (B.11) we have 
\begin{equation} 
\Delta\left(\mathrm{D}\left(\{(\Phi_{11}+3\Lambda),\Phi_{22},\Phi_{21}\} 
\right)\right)^\circ=0,  \label{y1} 
\end{equation} 
and by (NP.7) - (NP.9) 
\begin{equation} 
\Delta\left(\mathrm{D}\left(\{\lambda,\mu,\nu\}\right)\right)^\circ=0. 
\label{y2} 
\end{equation} 
  
\medskip 
  
We can now proceed inductively to show that all the $\mathrm{D}$-derivatives 
of $(\gamma+\bar\gamma), \tau, \rho, \sigma$ are $u$-independent on $% 
\widetilde{\mathcal{N}}_i$ up to order $n-1$ precisely when the functions $% 
\Phi_{00},\Phi_{01},\Phi_{02},(\Phi_{11}-3\Lambda)$ and $\mathrm{D}% 
(\Phi_{02})$ and their $\mathrm{D}$-derivatives up to order $n-2$ 
are $u$% 
-independent there. Suppose, as our inductive assumption, that the above 
statement is satisfied for $n=\bar n$. Then, by an exactly the 
same type of argument as the one yielded the equations (\ref{y1}) and (\ref 
{y2}) it  can be proven that 
\begin{equation} 
\Delta\left(\mathrm{D}^{(i)}(\{\lambda,\mu,\nu,(\Phi_{11}+3\Lambda),% 
\Phi_{22}, \Phi_{21}\})\right)^\circ=0, 
\end{equation} 
for any value of $i\in\{1,2,...,\bar n\}$. We show now that our inductive 
assumption holds also for $n=\bar n+1$. 
  
 First  note   that  by   (NP.6)  and   our  inductive  hypothesis 
$\Delta\left(\mathrm{D                                   }^{(\bar 
n)}(\gamma+\bar\gamma)\right)^\circ=0$   is   equivalent   to 
$\Delta\left(\mathrm{D}^{(\bar 
n-1)}(\Psi_2+\bar\Psi_2-2\Lambda+2\Phi_{11})     \right)^\circ=0$, 
which  in  virtue  of  (NP.12)  is  equivalent to the vanishing of 
$\Delta\left(\mathrm{D}^{(\bar 
n-1)}\left(\Phi_{11}\right)\right)^\circ$.  This, however, follows 
from  our  inductive  assumption,  i.e.  from the vanishing of the 
functions                           $\Delta\left(\mathrm{D}^{(\bar 
n-1)}\left(\Phi_{11}+3\Lambda\right)\right)$                   and 
$\Delta\left(\mathrm{D}                                   ^{(\bar 
n-1)}\left(\Phi_{11}-3\Lambda\right)\right)$   on    $\widetilde{ 
\mathcal{N}}_i$. 
  
By (NP.3) - (NP.5) we have that $\Delta\left(\mathrm{D}^{(\bar n)}(\{ 
\alpha,\beta,\tau\})\right)^\circ=0$ if and only if $\Delta\left(\mathrm{D}% 
^{(\bar n-1)}\left(\{\Phi_{01},\Psi_{1}\}\right) \right)^\circ=0$. The first 
part, i.e. $\Delta\left(\mathrm{D}^{(\bar 
n-1)}\left(\Phi_{01}\right)\right)^\circ=0$, follows from our inductive 
hypothesis while to see that $\Delta\left(\mathrm{D}^{(\bar 
n-1)}\left(\Psi_{1}\right)\right)^\circ=0$ consider the ($\bar n-2$)-times 
D-derivatives of (B.1). Using the commutators of D, $\delta$ and $\bar\delta$ 
we get that $\mathrm{D}^{(\bar n-1)}\left(\Psi_{1}\right)$ can be given in 
terms of $u$-independent quantities on $\widetilde{\mathcal{N}}_i$. 
  
Similarly, it follows from (NP.1) and (NP.2) that $\Delta\left(\mathrm{D}
^{(\bar n)}(\{\rho,\sigma\})\right)^\circ=0$ holds if and only if $\Delta\left(
\mathrm{D}^{(\bar n-1)}\left(\{\Phi_{00},\Psi_{0}\}\right)\right) ^\circ=0$. 
Again, the first half is just a part of inductive hypothesis while the 
vanishing of $\Delta\left(\mathrm{D}^{(\bar 
n-1)}\left(\Psi_{0}\right)\right) ^\circ$ follows from the fact that on $% 
\widetilde{\mathcal{N}}_i$ $\Delta\left(\mathrm{D}^{(\bar 
n-1)}\left(\Psi_{0}\right)\right)$ has to satisfy the equation 
\begin{equation} 
\Delta\left(\mathrm{D}^{(\bar n-1)}(\Psi_{0})\right) +\kappa_\circ\{ 
const\}\cdot\mathrm{D}^{(\bar n-1)}(\Psi_{0})+ \{terms\; independent\; of\; 
u\}=0 
\end{equation} 
which is yielded by making use of the ($\bar n-1$)-times D-derivative of 
(B.2) along with the application of the relevant form of the
commutators of D, $\delta$ and $\bar\delta$ several times. 
  
Consequently, the spin coefficients $(\gamma+\bar\gamma), \tau, \rho, \sigma$ 
and their $\mathrm{D}$-derivatives up to order $\bar n$ are independent of $u 
$ on $\widetilde{\mathcal{N}}_i$ which completes the proof of our inductive 
argument.\hfill \fbox{} 
  
\medskip 
  
 By  making  use  of  the  relation  between  the  Gaussian   null coordinates and the 
above applied null  tetrads, along with the relationship between the 
components of the energy-momentum tensor, $T_{ab}$, and the Ricci spinor  components 
$\Phi_{\alpha\beta}$ and $\Lambda$, Prop. \ref{prop-npind} can be rephrased as:
  
\begin{corollary} 
\label{cor-diff}  Let  $(\mathcal{O}_i,g_{ab}\vert_{\mathcal{O}_i})$ be 
an  elementary  spacetime  region.   Suppose  that  the components 
$T_{ur},T_{rr},T_{rA},T_{AB}$ along with the $r$-derivatives  of 
$T_{rr},T_{rA},T_{AB}$ up  to order  $n-1$ are  $u$-independent on 
$\widetilde{\mathcal{N}}_i$.   Then   the  $r$-derivatives   of  the 
functions  $ f,h_{A}$  and  $g_{AB}$  up  to  order $n$ are also 
$u$-independent on $\widetilde{\mathcal{N}}_i$. 
\end{corollary} 
 
\section{Particular Einstein-matter systems}\label{sec-mf} 
\setcounter{equation}{0} 
  
This section is  to introduce particular gravity-matter systems, 
such as EKG, E[nA]H, E[M]YMd and EYMH configurations, into our analysis. 
It is  important to emphasize that the dominant energy condition is 
not needed to be imposed separately because it follows from the  
particular form of the relevant energy momentum tensors that 
${T^a}_bk^b$ is proportional to $k^a$ on 
$\widetilde{\mathcal{N}}_i$ for these systems. 
  
\subsection{Einstein--Klein-Gordon--Higgs systems}\label{ssec-kg} 
  
A Klein-Gordon field is represented by a single real scalar field 
$\psi$ satisfying the linear second order hyperbolic equation 
\begin{equation} 
\nabla^a\nabla_a\psi+m^2\psi=0 \label{KG1} 
\end{equation} 
and the energy-momentum tensor of the relevant EKG system reads as 
\begin{equation} 
T_{ab}=\left(\nabla_a\psi\right)\left(\nabla_b\psi\right) 
-\frac{1}{2}g_{ab}\left[(\nabla^e\psi)(\nabla_e\psi)-m^2\psi^2\right] 
.\label{kgt} 
\end{equation} 
  
\begin{proposition}\label{prop-diff} 
Let  $(\mathcal{O}_i,g_{ab}\vert_{\mathcal{O}_i})$  be  an   elementary spacetime region  
associated with an EKG system. Then the functions $f,h_{A},g_{AB}$  and 
$\psi$, along  with  their  $r$-derivatives  up  to any order, are 
$u$-independent on $\widetilde{\mathcal{N}}_i$. 
\end{proposition} 
  
\noindent{\bf Proof}{\ } 
We shall prove our statement by induction. To see that the functions 
$f^\circ,h_{A}^\circ,g_{AB}^\circ$ and $\psi^\circ$ are $u$-independent note 
that as a consequence of (\ref{kgt}) $T_{uu}\geq 0$ on 
$\widetilde{\mathcal{N}}_i$ which, along with the argument applied in 
Prop. \ref{prop-3.1}, implies that $T_{uu}^\circ=0$ and, in turn, 
\begin{equation} 
\left( {\partial\psi\over \partial u}\right) 
^\circ=\left({\partial g_{AB}\over \partial u}\right) 
^\circ=0. \label{alpha} 
\end{equation} 
(\ref{kgt}) and (\ref{alpha}), implies then that 
\begin{equation} 
T_{uA}^\circ=0,\label{dec00} 
\end{equation} 
which by the argument of Prop. \ref{prop-4.1} yields that 
\begin{equation} 
\left( {\partial f\over \partial u}\right) 
^\circ=\left({\partial h_{A}\over \partial u}\right) 
^\circ=0. \label{beta} 
\end{equation} 
  
 To show  that the  first order  $r$-derivatives of  the functions 
$f,h_A$      and      $g_{AB}$      are      $u$-independent    on 
$\widetilde{\mathcal{N}}_i$,  in  virtue  of  Cor. \ref{cor-diff}, we 
need   to   demonstrate   that   $T_{ur},T_{rr},T_{rA},T_{AB}$ are 
$u$-independent  there.    Recall  now   that,  by   the  relation 
(\ref{kgt}),  the  components  $T_{ab}$  can  be given in terms of 
$f,h_A,g_{AB},\psi$  and the first order partial derivatives, 
$\partial\psi/\partial  x^\delta$, of $\psi$.    It 
follows  from  (\ref{alpha})  and  (\ref{beta})  that all of these 
functions  but  $\partial\psi/\partial  r$  are $u$-independent on 
$\widetilde{\mathcal{N}}_i$.      The     $u$-independentness     of 
$\partial\psi/\partial  r$  can  be  justified  as  follows: Since 
$\psi$ satisfies (\ref{KG1}) we  have that, in an  arbitrary local 
coordinate system $(x^1,x^2,x^3,x^4)$, 
\begin{equation} 
g^{\alpha\beta}\frac{\partial^2\psi}{\partial x^\alpha\partial x^\beta}- 
g^{\alpha\beta}{\Gamma^\delta}_{\alpha\beta} 
\frac{\partial\psi}{\partial x^\delta}+m^2\psi=0\label{m111} 
\end{equation} 
holds. 
In Gaussian null coordinates, i.e. whenever $x^1=u,x^2=r$, 
this equation reads on $\widetilde{\mathcal{N}}_i$ as 
\begin{equation} 
\frac{\partial^2\psi}{\partial u\partial r}+\kappa_\circ 
\frac{\partial\psi}{\partial r}+\{terms\; independent\; of\; u\}=0. 
\label{m1112} 
\end{equation} 
The $u$-derivative of (\ref{m1112}) is a homogeneous linear ordinary 
differential equation for ${\partial^2\psi}/{\partial u\partial r}$ along the 
generators of $\widetilde{\mathcal{N}}_i$ with the only $u$-periodic 
solution 
$(\partial^2\psi/{\partial u\partial r})^\circ\equiv 0$. Thus the first order 
$r$-derivatives of the functions $f,h_{A},g_{AB}$ and $\psi$ are 
$u$-independent on $\widetilde{\mathcal{N}}_i$. 
  
\medskip 
  
 Assume, now, as our inductive hypothesis that the $r$-derivatives 
of the functions  $f,h_{A},g_{AB}$ and $\psi$  are $u$-independent 
on $\widetilde{\mathcal{N}}_i$  up to  order $\bar  n\in\mathbb{N}$. 
We   need   to   show   that   $r$-derivatives   of  the functions 
$f,h_{A},g_{AB}$  and  $\psi$  up  to  order  $\bar  n+1$ are also 
independent  of  $u$  on  $\widetilde{\mathcal{N}}_i$.  In virtue of 
Cor. \ref{cor-diff}, the $u$-independentness of the $r$-derivatives 
$f,h_{A}$ and $g_{AB}$ up to  order $\bar n+1$ can be  traced back 
to   the   $u$-independentness    of   the   $r$-derivatives    of 
$T_{rr},T_{rA}$  and  $T_{AB}$  up  to  order  $\bar n$.  However, 
according to (\ref{kgt}), these derivatives can be given in  terms 
of   the   $r$-derivatives   of   $\psi$,   $\partial\psi/\partial 
x^\alpha$, and $f,h_A,g_{AB}$ up to order $\bar n$.  Note that, by 
our    inductive    assumption,     we    have    that     $\psi$, 
$\partial\psi/\partial  u$,  $\partial\psi/\partial  x^A$  and the 
functions $f,h_A, g_{AB}$ possess $u$-independent  $r$-derivatives 
up  to  order  $\bar  n$.  Thereby, only the $u$-independentness of the 
$r$-derivatives of $\partial\psi/\partial r$ up to the same order
have to be shown. To do this, differentiate first (\ref{m111}) $\bar n$-times 
with respect to $r$ and set $r=0$. The yielded equation is
\begin{equation} 
\frac{\partial^{\bar n+2}\psi}{\partial u\partial^{\bar n+1} r}+ 
2\kappa_\circ\frac{\partial^{\bar n+1}\psi}{\partial^{\bar n+1} r}+ 
\{terms\; independent\; of\; u\}=0.\label{mom} 
\end{equation} 
Then, by differentiating (\ref{mom}) with respect to $u$, we get 
a homogeneous linear ordinary differential equation for 
$(\partial^{\bar n+2}\psi/\partial u\partial r^{\bar n+1})$  with the only 
$u$-periodic solution $(\partial^{\bar n+2}\psi/\partial u\partial r^{\bar n+1})^\circ 
\equiv 0$. \hfill \fbox{} 
  
\begin{remark} 
Essentially the same simple reasoning does apply to the case of a set of self-interacting 
complex  scalar fields, whenever the interaction terms do not contain  derivatives of the fields. 
 The above  argument also  generalize to  the following  case of 
[non-Abelian] Higgs  fields: Let  $\mathfrak{g}$ be  a Lie  algebra 
associated  with  a  Lie  group  $\mathrm{G}$.   For  the  sake of 
definiteness $\mathrm{G}$ will be  assumed 
to be a matrix group and it will also be assumed that there exists 
a positive definite real inner  product, denoted by $(\ /\  )$, on 
$\mathfrak{g}$   which    is   invariant    under   the    adjoint 
representation.   (The  relevant  gauge or matrix indices will be  suppressed throughout.) 
Denote by $\psi:\mathcal{O}_i\rightarrow  \mathfrak{g}$ the Higgs field and consider 
the associated  matter Lagrangian
\begin{equation} 
\mathcal{L}_{_{\mathrm{matter}}}^{^{_{Higgs}}}=2\left[g^{ab} 
\left(\nabla_a\psi/\nabla_b\psi\right)-V\left(\psi\right)\right], 
\end{equation} 
where  $V$  is  a  sufficiently  regular but otherwise arbitrary gauge invariant expression 
of the field variable $\psi$. The claim that  Prop. \ref{prop-diff} generalizes  to the 
corresponding E[nA]H systems is based on the following observations: First, since the real 
inner  product, $(\ /\  )$, is  positive  definite, by a 
straightforward  modification  of  the  first  part of the proof of 
Prop. 5.1,   the vanishing of $T^{\circ}_{uu}$ and $T^{\circ}_{uA}$ can be 
justified for an adapted gauge representation $\psi$.  Second, because  of  
the  `special'  choice  of  the interaction term,  the equations  for the  $r$-derivatives 
of  the various gauge components of $\psi$ decouple so that each  equation possesses the 
fundamental form of (\ref{m1112}) or (\ref{mom}). This way we get the following:
\end{remark} 
  
\begin{corollary}\label{cor-diff2} 
Let  $(\mathcal{O}_i,g_{ab}\vert_{\mathcal{O}_i})$  be  an   elementary 
spacetime  region associated with an E[nA]H system as it was specified above.    
Then there exists a gauge representation $\psi$ so that the functions 
$f,h_{A},g_{AB}$ and $\psi$,  along with their  $r$-derivatives up 
to any order, are $u$-independent on $\widetilde{\mathcal{N}}_i$. 
\end{corollary} 
  
\subsection{Einstein--Maxwell systems}\label{ssec-mf} 
  
 In this  subsection we  shall consider  the case  of source  free electromagnetic fields.  
The proof of Prop. \ref{prop-diffmf} is an  alternative  of  the  argument  given   in 
\cite{im1}  and  it  is  presented  to  provide a certain level of 
preparation for the more complicated case of EYM systems. 
  
 A  source  free  electromagnetic  field,  in any simply connected 
elementary  spacetime  region,  can  be  represented  by  a vector 
potential $A_a$ related to the Maxwell tensor as 
\begin{equation} 
F_{ab}=\nabla_aA_b-\nabla_bA_a. 
\end{equation} 
The energy-momentum tensor of the related Einstein-Maxwell 
system reads as 
\begin{equation} 
T_{ab}=-{1\over 4\pi} \left\{F_{ae}{F_{b}}^{e} 
-\frac{1}{4}g_{ab}\left(F_{ef}F^{ef}\right)\right\}\label{mt} 
\end{equation} 
while the Maxwell equations are 
\begin{equation} 
\nabla^aF_{ab}=0.\label{me} 
\end{equation} 
  
\begin{proposition}\label{prop-diffmf} 
Let  $(\mathcal{O}_i,g_{ab}\vert_{\mathcal{O}_i})$  be  an   elementary 
spacetime region and consider a source free electromagnetic  field 
$F_{ab}$ on $\mathcal{O}_i$ in Einstein theory.  Then there exists a vector potential 
$A_a$  associated  with  $F_{ab}$   so  that  $A_u$  vanishes   on 
$\widetilde{\mathcal{N}}_i$,  moreover,  the  $r$-derivatives of the 
functions  $f,h_A$  and  $g_{AB}$  and  that  of  the   components 
$A_u,A_r,A_B$   up   to   any   order   are   $u$-independent   on 
$\widetilde{\mathcal{N}}_i$. 
\end{proposition} 
  
\noindent{\bf Proof}{\ } 
 We shall prove  the above statement  by induction.  To  start off 
pick up  an arbitrary,  $u$-periodic, vector  potential $A_a'$  of 
$F_{ab}$.  We have the  `gauge freedom'  of 
adding the  gradient $\nabla_a\alpha$  of a  function $\alpha$  to 
$A_a'$. At least the major part of this freedom can  be fixed by the specification  of 
the gauge  source function  ${\mathcal A'}=\nabla^aA_{a}'$. Since the components
of $g_{ab}$ and $A_a'$ were assumed to be $u$-periodic 
${\mathcal A'}$ is also a smooth $u$-periodic function on $\mathcal{O}_i$. 

\begin{lemma}\label{maxl} 
There exists a smooth $u$-periodic function 
$\alpha:\mathcal{O}_i\rightarrow\mathbb{R}$ such that 
\begin{equation} 
\partial\alpha/\partial u=-A_u' \label{gauge1} 
\end{equation} 
holds on $\widetilde{\mathcal{N}}_i$, moreover, the gauge source function 
\begin{equation} 
{\mathcal A}:=\nabla^a\nabla_a\alpha+{\mathcal A'}=0\label{gsf} 
\end{equation} 
on $\mathcal{O}_i$. 
\end{lemma} 
  
\noindent{\bf Proof}{\ } 
To start off let $\alpha_{_{(0)}}$ be a solution of
(\ref{gauge1}) on $\widetilde{\mathcal{N}}_i$, 
i.e. $\alpha_{_{(0)}}=-\int_{u_o}^u A_u'^\circ + \chi$ where $\chi$ is 
an arbitrarily chosen smooth function of $x^3$ and $x^4$. In virtue of 
(\ref{fub}) $F_{uB}^\circ=\left({\partial A_B'/\partial u} 
-{\partial A_u'/\partial x^B}\right)^\circ=0$ (the vanishing of $F_{uB}^\circ$ 
is independent of the applied vector potential) we have that 
\begin{equation} 
{\partial\alpha_{_{(0)}}\over \partial x^B} = - A_B'^\circ+\chi^*, 
\label{alpha02} 
\end{equation} 
where $\chi^*$ is a smooth function of $x^3$ and $x^4$. Since the
r.h.s. of (\ref{alpha02}) is 
$u$-periodic $\alpha_{_{(0)}}$ has also to be $u$-periodic. 
  
To show that there exists a smooth function $\alpha$ on $\mathcal{O}_i$ 
so that both 
(\ref{gauge1}) and (\ref{gsf}) are satisfied we shall use the
characteristic initial value problem associated with (\ref{gsf}). 
Notice first that (\ref{gauge1}) will be immediately satisfied if the
relevant solution of (\ref{gsf}) is ensured to coincide with $\alpha_{_{(0)}}$
on $\widetilde{\mathcal{N}}_i$. 
To specify our initial data hypersurface consider first a smooth
cross-section $\widetilde{\sigma}_i(u_\circ)$ [$u=u_\circ\in\mathbb{R}$] 
of $\widetilde{\mathcal{N}}_i$. This cross-section divides the
boundary of the causal future $J^+[\widetilde{\sigma}_i(u_\circ),
\mathcal{O}_i]$ of $\widetilde{\sigma}_i(u_\circ)$ in $\mathcal{O}_i$
into two connected pieces. Denote by $\widetilde{\mathcal{N}}_i^+(u_\circ)$ 
the component contained by $\widetilde{\mathcal{N}}_i$ and by 
$\widetilde{\mathcal{N}}_i^-(u_\circ)$ the other part. Since there is
no conjugate point to $\widetilde{\sigma}_i(u_\circ)$ along the null
generators of the smooth hypersurfaces $\widetilde{\mathcal{N}}_i^+(u_\circ)$ 
and $\widetilde{\mathcal{N}}_i^-(u_\circ)$ they comprise a suitable
initial data surface $\widetilde{\Sigma}_i(u_\circ)=
\widetilde{\mathcal{N}}_i^+(u_\circ)\cup \widetilde{\mathcal{N}}_i^-(u_\circ)$
for (\ref{gsf}). To specify the initial data extend first the function 
$\alpha_{_{(0)}}$ from $\widetilde{\mathcal{N}}_i$ onto $\mathcal{O}_i$ 
by keeping its value to be constant along the integral curves of 
$l^a=(\partial/\partial r)^a$. Choose then the restriction of this
extension $\alpha_{_{(0)}}$ onto $\widetilde{\Sigma}_i(u_\circ)$ as
our initial data. 

Consider now the discrete isometry action $\Psi_i^{(k)}:\mathcal{O}_i
\rightarrow \mathcal{O}_i$ [$k\in \mathbb{Z}$], associated with the
fiber preserving local isometry action  $\psi_i$ (see section 2). By
construction for any
value of $k\in \mathbb{Z}$ $\Psi_i^{(k)}$ maps $\widetilde{\Sigma}_i(u_\circ)$
to the surface $\widetilde{\Sigma}_i(u_\circ+kP)$. Moreover, since  
$\alpha_{_{(0)}}$ is invariant under the action of $\Psi_i^{(k)}$, 
the initial data specifications 
$\alpha_{_{(0)}}\vert_{\widetilde{\Sigma}_i(u_\circ)}$ and 
$\alpha_{_{(0)}}\vert_{\widetilde{\Sigma}_i(u_\circ+kP)}$ are mapped
onto each other by $\Psi_i^{(k)}$. Appealing now to the uniqueness of
the solutions to the linear wave equation (\ref{gsf}) in the domain of
dependence of an initial data hypersurface it follows then that the 
relevant Cauchy developments of 
$\alpha_{_{(0)}}\vert_{\widetilde{\Sigma}_i(u_\circ)}$ and 
$\alpha_{_{(0)}}\vert_{\widetilde{\Sigma}_i(u_\circ+kP)}$ are also
mapped onto each other by $\Psi_i^{(k)}$. This, however, in turn
implies that the unique smooth solution $\alpha$ of (\ref{gsf}) with
initial data $\alpha_{_{(0)}}\vert_{\widetilde{\Sigma}_i(u_\circ)}$ is
$u$-periodic with periodicity length $P$. By making use of this
property $\alpha$ can be extended onto the entire of 
$J^+[\widetilde{\mathcal{N}}_i]\cap\mathcal{O}_i$ so that the
extension coincides with $\alpha_{_{(0)}}$ on $\widetilde{\mathcal{N}}_i$. 

An analogous argument applies to the complementary region, 
$J^-[\widetilde{\mathcal{N}}_i,\mathcal{O}_i]$, which completes then
the proof of the second part\footnote{% 
I would like to say thank to the unknown referee who pointed out the
incompleteness of the relevant argument contained by the former
version of this paper.} of our lemma.\hfill \fbox{} 
  
\bigskip 
  
Turning back to the proof of Prop. 5.2 consider 
the vector potential $A_a:=A_a'+\nabla_a\alpha$. 
In virtue of (\ref{gauge1}), we have that 
\begin{equation} 
A_u^\circ=0.\label{au0} 
\end{equation} 
Moreover, we also have that ${\mathcal A}=\nabla^aA_{a}$
vanishes in $\mathcal{O}_i$. 
  
From this point one can proceed as follows: By (\ref{mt}) and the negative 
definiteness of $g_{AB}$ one gets that $T_{uu}^\circ=-{1\over 4\pi}(F_{uA} 
F_{uB}g^{AB})^\circ\geq0$ which, along with the argument 
of Prop. \ref{prop-3.1}, implies that 
\begin{equation} 
F_{uB}^\circ=\left({\partial A_B\over \partial u} 
-{\partial A_u\over\partial x^B}\right)^\circ=0 \ \ \ {\rm and} \ \ 
\ \left({\partial g_{AB}\over \partial u}\right)^\circ=0.\label{fub} 
\end{equation} 
In virtue of (\ref{au0}) and (\ref{fub}) we also have that 
\begin{equation} 
\left(\frac{\partial A_B }{\partial u}\right)^\circ=0.\label{fuB} 
\end{equation} 
Then (\ref{fub}) and (\ref{mt}) imply that $T_{uA}^\circ=0$ which by the 
argument of Prop. \ref{prop-4.1} yields that 
\begin{equation} 
\left( {\partial f\over \partial u}\right) 
^\circ=\left({\partial h_{A}\over \partial u}\right) 
^\circ=0. \label{beta2} 
\end{equation} 
What remains to be shown, to complete the first step of our inductive 
proof, is the $u$-independentness of $A_r$ on 
$\widetilde{\mathcal{N}}_i$ which can be demonstrated as follows: In 
Gaussian null coordinates the `$u$'-component of the Maxwell equation 
(\ref{me}) on $\widetilde{\mathcal{N}}_i$ takes the form 
\begin{equation} 
\frac{\partial F_{ru} }{\partial u}=0,\label{fur0} 
\end{equation} 
i.e., 
\begin{equation} 
\frac{\partial }{\partial u}\left(\frac{\partial A_r }{\partial u}- 
\frac{\partial A_u }{\partial r}\right)^\circ=0.\label{ru} 
\end{equation} 
By differentiating now $\nabla^aA_{a}=0$ with respect to $u$ and 
setting $r=0$ we get 
\begin{equation} 
\frac{\partial }{\partial u}\left(\frac{\partial A_r }{\partial u}+ 
\frac{\partial A_u }{\partial r}\right) 
+2\kappa_\circ\frac{\partial A_r }{\partial u}=0.\label{i} 
\end{equation} 
It follows from (\ref{ru}) and (\ref{i}) that 
\begin{equation} 
\frac{\partial }{\partial u}\left(\frac{\partial A_r }{\partial u}\right) 
+\kappa_\circ\frac{\partial A_r }{\partial u}=0\label{0} 
\end{equation} 
holds on $\widetilde{\mathcal{N}}_i$, which, in turn, along 
with periodicity, implies then that $A_r$ has to be $u$-independent on 
$\widetilde{\mathcal{N}}_i$ as we wanted to demonstrate. 
  
\medskip 
  
To see that the first $r$-derivative of the functions $f,h_A,g_{AB}$, 
moreover, that of the components $A_u,A_r,A_B$ are $u$-independent on 
$\widetilde{\mathcal{N}}_i$ we can proceed as follows: Note first that 
by the $u$-independentness of $A_r$ on $\widetilde{\mathcal{N}}_i$ and 
by (\ref{ru}) we also have that $(\partial A_u/\partial r)^\circ$ is 
$u$-independent. 
  
The $u$-independentness of $(\partial A_B/\partial r)^\circ$ can be 
shown by making use of the `$B$'-component of the Maxwell equation 
(\ref{me}) which reads on $\widetilde{\mathcal{N}}_i$ as 
\begin{equation} 
\frac{\partial}{\partial u}\left(\frac{\partial A_B}{\partial r}\right)+ 
\kappa_\circ\frac{\partial A_{B} }{\partial r}+ 
\{terms\; independent\; of\; u\}=0 \label{dab} 
\end{equation} 
with the only periodic solution 
$(\partial^{2} A_B/\partial u\partial r)^\circ\equiv 0$. It follows 
then that all the components of $F_{ab}$ are $u$-independent on 
$\widetilde{\mathcal{N}}_i$ which, along with (\ref{mt}) and the 
$u$-independentness of the functions $f,h_A$ and $g_{AB}$, implies 
that each of the components $T_{ur},T_{rr},T_{rA}$ and $T_{AB}$ is 
$u$-independent on 
$\widetilde{\mathcal{N}}_i$. This later property, however, in virtue of 
Cor. \ref{cor-diff} yields that the first $r$-derivatives of the functions 
$f,h_A$ and $g_{AB}$ are $u$-independent on $\widetilde{\mathcal{N}}_i$. 
Thus only the $u$-independentness of $(\partial A_r/\partial r)^\circ$ 
remained to be shown which can be done as follows: The first 
$r$-derivative of the `$u$'-component of the Maxwell equation 
(\ref{me}) gives that on $\widetilde{\mathcal{N}}_i$ 
\begin{equation} 
\frac{\partial }{\partial u}\left(\frac{\partial F_{ur} }{\partial r}\right)+ 
\{terms\; independent\; of\; u\}=0 \label{dfur} 
\end{equation} 
holds which, along with the $u$-periodicity, implies that 
\begin{equation} 
\frac{\partial }{\partial u}\left[\frac{\partial }{\partial r} 
\left(\frac{\partial A_r }{\partial u}- 
\frac{\partial A_u }{\partial r}\right)\right]^\circ=0.\label{dru} 
\end{equation} 
Moreover, the vanishing of the first $r$-derivative of the gauge 
source function $\mathcal A$ on $\widetilde{\mathcal{N}}_i$ gives that 
\begin{equation} 
\frac{\partial }{\partial u}\left[\frac{\partial }{\partial r} 
\left(\frac{\partial A_r }{\partial u}+\frac{\partial A_u }{\partial r}\right)+ 
4\kappa_\circ \frac{\partial A_r }{\partial r}\right]^\circ=0. 
\label{dru2} 
\end{equation} 
In virtue of (\ref{dru}) and (\ref{dru2}) we have that both 
$(\partial^2 A_r/\partial u\partial r)^\circ$ and 
$(\partial^3 A_r/\partial u\partial r^2)^\circ$ vanish identically. 
  
\medskip 
  
Assume, now, as our inductive hypothesis that the $r$-derivatives of 
the functions $f,h_A,g_{AB}$ and that of $A_u,A_r,A_B$ are 
$u$-independent on $\widetilde{\mathcal{N}}_i$ up to order $\bar n\in 
\mathbb{N}$. Then, by the vanishing of the $\bar n^{th}$ $r$-derivative 
of $\mathcal A$ together with the $u$-independentness of 
$(\partial^{\bar n} A_r/\partial r^{\bar n})^\circ$ on 
$\widetilde{\mathcal{N}}_i$, we also have that 
\begin{equation} 
\frac{\partial }{\partial u}\left(\frac{\partial^{\bar n+1} A_u } 
{\partial r^{\bar n+1}}\right)^\circ=0.\label{dnp1} 
\end{equation} 
Then the $u$-independentness of 
$(\partial^{\bar n+1} A_B/\partial r^{\bar n+1})^\circ$ can be seen as 
follows: Differentiate the `$B$'-component of the Maxwell equation 
(\ref{me}) $\bar n$-times with respect to $r$ and set $r=0$. This way 
we get 
\begin{equation} 
\frac{\partial }{\partial u}\left(\frac{\partial^{\bar n+1}A_B} 
{\partial r^{\bar n+1}}\right)+ 
2\kappa_\circ\frac{\partial^{\bar n+1} A_B }{\partial r^{\bar n+1}}+ 
\{terms\; independent\; of\; u\}=0 \label{dab2} 
\end{equation} 
which implies, along with the $u$-periodicity, that 
$(\partial^{\bar n+2} A_B/\partial u\partial r^{\bar n+1})^\circ\equiv 0$. 
It follows then that the $r$-derivatives of the components of $F_{ab}$ 
up to order $\bar n$ are $u$-independent on 
$\widetilde{\mathcal{N}}_i$ which, in turn, along with our inductive hypotheses, 
implies that the $r$-derivatives of the components $T_{rr},T_{rA}$ and 
$T_{AB}$ are $u$-independent there. This later property, however, in virtue of 
Cor.\ref{cor-diff}, guarantees the $u$-independentness of the $r$-derivatives 
$f,h_{A}$ and $g_{AB}$ up to order $\bar n+1$. Thus to complete our inductive 
proof we need only to demonstrate the $u$-independentness of 
$(\partial^{\bar n+1} A_r/\partial r^{\bar n+1})^\circ$ which can be done as 
follows: Differentiate $(\bar n+1)$-times the `$u$'-component of (\ref{me}) 
with respect to $r$ and set $r=0$. The yielded equation is of the form 
\begin{equation} 
\frac{\partial }{\partial u}\left(\frac{\partial^{\bar n+1}F_{ur}} 
{\partial r^{\bar n+1}}\right)+ 
\{terms\; independent\; of\; u\}=0, \label{dab3} 
\end{equation} 
which, along with the $u$-periodicity, implies that 
\begin{equation} 
\frac{\partial }{\partial u}\left[\frac{\partial^{\bar n+1} }{\partial 
r^{\bar n+1}} \left(\frac{\partial A_r }{\partial u}- 
\frac{\partial A_u }{\partial r}\right)\right]^\circ=0.\label{dru11} 
\end{equation} 
In addition, the vanishing of the $(\bar n+1)^{th}$ $r$-derivative of the gauge 
source function $\mathcal A$ on $\widetilde{\mathcal{N}}_i$ yields that 
\begin{equation} 
\frac{\partial }{\partial u}\left[\frac{\partial^{\bar n+1}} 
{\partial r^{\bar n+1}} 
\left(\frac{\partial A_r }{\partial u}+\frac{\partial A_u }{\partial r}\right)+ 
4\kappa_\circ \frac{\partial^{\bar n+1} A_r}{\partial r^{\bar n+1}}\right]^\circ=0. 
\label{dru12} 
\end{equation} 
In virtue of the last two equations, (\ref{dru11}) and (\ref{dru12}), we have 
that both $(\partial^{\bar n+2} A_r/\partial u\partial r^{\bar n+1})^\circ$ and 
$(\partial^{\bar n+3} A_r/\partial u\partial r^{\bar n+2})^\circ$
vanish, as we wanted to demonstrate.\hfill \fbox{} 
  
\subsection{Einstein--[Maxwell]--Yang-Mills (--dilaton --Higgs) systems}
\label{ssec-ym} 
  
Let us consider now the case of a YM gauge field which can be represented by a vector 
potential $A_a$ taking values in the Lie algebra $\mathfrak{g}$ of a matrix group $\mathrm{G}$,
i.e. $\mathrm{G}\subset\mathrm{GL}(N,\mathbb{C})$ for some $N\in\mathbb{N}$.
In terms of the vector potential $A_a$ the Lie-algebra-valued 2-form field $F_{ab}$ is given as 
\begin{equation} 
F_{ab}=\nabla_aA_b-\nabla_bA_a+ 
\left[A_a,A_b\right]\label{ymf} 
\end{equation} 
where $[\ ,\ ]$ denotes the product in $\mathfrak{g}$. 
The energy-momentum tensor of the related EYM system is 
\begin{equation} 
T_{ab}=-{1\over 4\pi}\left\{\left(F_{ae}/{F_{b}}^{e}\right) 
-\frac{1}{4}g_{ab}\left(F_{ef}/F^{ef}\right)\right\},\label{ymt} 
\end{equation} 
where $(\ /\ )$ is a positive definite real inner product in $\mathfrak{g}$ 
which is invariant under the adjoint representation. 
Finally, the field equations of such a YM field read as 
\begin{equation} 
\nabla^aF_{ab}+\left[A^a,F_{ab}\right]=0.\label{yme} 
\end{equation} 
  
We would like to generalize Prop. 5.2 for such an EYM system. In doing this start 
with an arbitrary $u$-periodic adapted gauge potential $A_a'$ defined on a simply connected 
elementary spacetime region ${\mathcal O}_i$. It is known that there is a freedom in representing 
a YM field, i.e. instead of $A_a'$ we can also use the gauge related field 
\begin{equation} 
A_a=m^{-1}\left(\nabla_a m+A_a'm\right),\label{ymgau} 
\end{equation} 
where $m:{\mathcal O}_i\rightarrow\mathrm{G}$ is an arbitrary smooth function. 
Then the following can be proven: 
  
\begin{lemma} 
There exists a smooth $u$-periodic function 
$m:\mathcal{O}_i\rightarrow\mathrm{G}$ such that 
\begin{equation} 
A_u^\circ=\left[m^{-1}\left(\partial m/\partial u+A_u'm\right)\right]^\circ 
:=a_\circ \label{ymgauge1} 
\end{equation} 
is a $u$-independent $\mathfrak{g}$-valued function on
$\widetilde{\mathcal{N}}_i$,  moreover, the gauge source function 
\begin{equation} 
{\mathcal A}:=\nabla^aA_a \left( =\nabla^a\left[m^{-1}\left(\nabla_a m+
 A_a'm\right)\right]\right) 
\label{ymgsf} 
\end{equation} 
vanishes on $\mathcal{O}_i$.  
\end{lemma}

\noindent{\bf Proof}{\ } Consider first the (matrix) differential equation 
\begin{equation} 
\partial m^*/\partial u+A_u'^\circ m^*=0 \label{ymgauge2} 
\end{equation} 
where $m^*$ is a $\mathrm{G}$-valued function on $\widetilde{\mathcal{N}}_i$. The general 
solution of (\ref{ymgauge2}) is of the form (see e.g. Theorem 2.2.5 of \cite{farkas}) 
\begin{equation} 
m^*=m_{_{(0)}}\mathrm{exp}(-a_\circ u) \label{ymgauge3} 
\end{equation} 
where $m_{_{(0)}}:\widetilde{\mathcal{N}}_i\rightarrow\mathrm{G}$ is a 
$u$-periodic function, with the same period as $A'_u$, and 
$a_\circ:\widetilde{\mathcal{N}}_i\rightarrow\mathfrak{g}$ is independent of $u$. 
Then, (\ref{ymgauge2}) and (\ref{ymgauge3}) implies that
\begin{equation} 
\partial m_{_{(0)}}/\partial u+A_u'^\circ m_{_{(0)}}-m_{_{(0)}}a_\circ=0. \label{ymgauge4} 
\end{equation} 
It follows then from (\ref{ymgau}) that for any choice of a gauge
transformation $m$ with $m\vert_{\widetilde{\mathcal{N}}_i}=
m_{_{(0)}}(u,x^3,x^4)$ 
$A_u^\circ=a_\circ$ holds on $\widetilde{\mathcal{N}}_i$.

The second part of our statement can be proven by making use of an
argument analogous to that of the proof of the second part of
lemma \ref{maxl}. The only significant difference is that the evolution
equation (\ref{gsf}) for $\alpha$ has to be
replaced by 
\begin{equation} 
\nabla^a\nabla_am+{\mathcal{A}'}m+ A_a'\nabla^am- 
(\nabla^am)m^{-1}(\nabla_am+ A_a'm)=0 \label{ymlem3} 
\end{equation} 
and the relevant initial data for $m$ has to be constructed by making
use of the above defined $m_{_{(0)}}$.\hfill \fbox{} 
 
\begin{remark}\label{rem-amb}
Hereafter, without loss of generality, we shall assume that the $u$-component of an adapted 
gauge potential is $u$-independent. Note, however, that neither $a_\circ$ nor its eigenvalues 
are uniquely determined by (\ref{ymgauge2}). To see this note that $m^*=m_{_{(0)}}\cdot
\widetilde m\cdot\mathrm{exp}[{2\pi iu}/{P}\cdot\mathrm{diag}(k_1,...,k_N)]
\cdot\mathrm{exp}(-\widetilde a_\circ u)$ is also a solution of (\ref{ymgauge2})  with 
\begin{equation}
\widetilde a_\circ=\widetilde m^{-1}a_\circ\widetilde m+{2\pi i}/{P}\cdot\mathrm{diag}(k_1,...,k_N),
\label{trgen}
\end{equation}
where $\widetilde m:\widetilde{\mathcal{N}}_i\rightarrow\mathrm{G}$ is $u$-independent and the
entries $k_i$ are integers so that the matrices $\widetilde m^{-1}a_\circ\widetilde m$ and 
$\mathrm{diag}(k_1,...,k_{N})$ commute. The simplest possible $u$-periodic gauge transformation 
$m:{\mathcal{O}}_i\rightarrow\mathrm{G}$ manifesting this freedom can be given as
\begin{equation}
m=\widetilde m(x^3,x^4)\cdot 
\mathrm{exp}\left[{2\pi iu}/{P}\cdot\mathrm{diag}(k_1,...,k_N)\right].\label{defm}
\end{equation}
\end{remark} 

\begin{remark}
In what follows key role will be played by the possible behavior of $u$-periodic 
$\mathfrak{g}$-valued functions satisfying differential equations -- see e.g. 
(\ref{yfuB1}), (\ref{yfur0}) and (\ref{y0}) -- of the form 
\begin{equation} 
\frac{\partial \mathcal{F} }{\partial u}=c_1\kappa_\circ \mathcal{F}+
c_2[a_\circ,\mathcal{F}],\label{ymgen} 
\end{equation} 
with $c_1=0,c_2=-1$ or $c_1=-1,c_2=-1/2$. This equation can also be read as 
\begin{equation} 
\frac{\partial \mathcal{F} }{\partial u}=\mathfrak{C}(a_\circ;\kappa_\circ,c_1,c_2) \mathcal{F},
\label{ymgen2} 
\end{equation}  
where now $\mathcal{F}$ is considered to be a `vector' whereas $\mathfrak{C}(a_\circ;
\kappa_\circ,c_1,c_2)$ is a linear map acting on the corresponding `vector space' given as 
\begin{equation}  
\mathfrak{C}(a_\circ;\kappa_\circ,c_1,c_2)=c_1\kappa_\circ \mathbb{E}\otimes\mathbb{E}+
c_2\left(a_\circ\otimes\mathbb{E}-\mathbb{E}\otimes a^t_\circ\right).\label{defC}
\end{equation}  
Here $\mathbb{E}$, $\otimes$ and $a^t_\circ$ denote the unit element of $G$, the tensor product 
on $G$ and the transpose of $a_\circ$, respectively. The system (\ref{ymgen2}) is known (see e.g. 
\cite{farkas}) to have a $u$-periodic solution of period $P$ if and only if $2\pi i k/P$ is an 
eigenvalue of $\mathfrak{C}(a_\circ;\kappa_\circ,c_1,c_2)$ for some $k\in\mathbb{Z}$. If $k=0$ 
then  $\mathfrak{C}(a_\circ;\kappa_\circ,c_1,c_2)$ is singular and (\ref{ymgen2}) has non-trivial 
$u$-independent solutions. If $k\in\mathbb{Z}\setminus\{0\}$ then (\ref{ymgen2}) has a $u$-periodic 
solutions with smallest positive period $P/\vert k\vert$. 
\end{remark}

\begin{definition} 
Let $A_a$ be an adapted gauge representation of a YM field with $A_u^\circ=a_\circ$. Then 
$A_a$ is called to be a `preferred' gauge representation if neither of the eigenvalues of 
$\mathfrak{C}(a_\circ;\kappa_\circ,c_1,c_2)$, with $c_1=0,c_2=-1$ or $c_1=-1,c_2=-1/2$, 
is of the form $2\pi i k/P$ for any $k\in\mathbb{Z}\setminus\{0\}$.
\end{definition} 

\begin{remark}\label{rem-amb2}
The eigenvalues of $\mathfrak{C}(a_\circ;\kappa_\circ,c_1,c_2)$ are the roots of its characteristic 
polynomial. The characteristic polynomial of $\mathfrak{C}(a_\circ;\kappa_\circ,c_1,c_2)$ is in fact 
an `invariant polynomial' because its coefficients can be given in terms of $\kappa_\circ,c_1,c_2$ 
and polynomials of the traces of various powers of $a_\circ$. In addition, 
(\ref{yfub}) and (\ref{yfuB}) imply\footnote{%
These equations are valid for any $u$-periodic gauge representation $A_a$ having $u$-independent
$A_u^\circ=a_\circ$.} that ${\partial a_\circ}/{\partial x^B}=-[a_\circ, A_B^\circ]$ 
thereby these traces have to be constant throughout $\widetilde{\mathcal{N}}_i$, i.e. the 
characteristic polynomial of $\mathfrak{C}(a_\circ;\kappa_\circ,c_1,c_2)$ is the simplest possible 
type  with degree $\mathrm{dim(G)}=N^2$ and with constant coefficients.
\end{remark} 

\begin{lemma}\label{yml} 
Let $A_a$ be an adapted gauge representation of a YM field with $A_u^\circ=a_\circ$. Then there 
exist $\widetilde m:\widetilde{\mathcal{N}}_i\rightarrow\mathrm{G}$ and $k_i\in\mathbb{Z}$ so 
that the gauge representation $\widetilde A_a$ yielded by (\ref{defm}) is preferred. 
\end{lemma}

\noindent{\bf Proof}{\ } 
Denote by $a_\circ^{_{J}}$ the Jordan normal form of $a_\circ$ and assume that the 
$\mathrm{G}$-valued function $m_{_{J}}$ is so that $a_\circ^{_{J}}=m_{_{J}}^{-1}a_\circ m_{_{J}}$.
Then it follows from $\mathfrak{C}(m^{-1}_{_{J}}a_\circ m_{_{J}};\kappa_\circ,c_1,c_2)=
(m_{_{J}}\otimes m_{_{J}})^{-1}\cdot\mathfrak{C}(a_\circ;\kappa_\circ,c_1,c_2)
\cdot(m_{_{J}}\otimes m_{_{J}})$ that the eigenvalues of $\mathfrak{C}(a_\circ;
\kappa_\circ,c_1,c_2)$ and $\mathfrak{C}(a_\circ^{_{J}};\kappa_\circ,c_1,c_2)$ coincide.
Denote by $\alpha_i^{_{J}}$ the eigenvalue of the $i^{th}$ ($i\in\{1,...,N_J\}$) 
Jordan block of $a_\circ^{_{J}}$. Using then the special block structure of $a_\circ^{_{J}}$, 
along with the definition (\ref{defC}), the eigenvalues of 
$\mathfrak{C}(a_\circ^{_{J}};\kappa_\circ,c_1,c_2)$ can be seen to be 
\begin{equation} 
\mathfrak{c}_{ij}=c_1\kappa_\circ+c_2(\alpha_i^{_{J}}-\alpha_j^{_{J}}),\label{eig}
\end{equation}
where for distinct eigenvalues $\alpha_i^{_{J}}$ and $\alpha_j^{_{J}}$ of $a_\circ^{_{J}}$ the 
multiplicity  of the eigenvalue $\mathfrak{c}_{ij}$ is $m(\mathfrak{c}_{ij})=m(\alpha_i^{_{J}})
m(\alpha_j^{_{J}})$ whereas the multiplicity of $\mathfrak{c}_{0}=c_1\kappa_\circ$ is 
$m(\mathfrak{c}_{0})=N^2-2\sum_{\alpha_i^{_{J}}\not=\alpha_j^{_{J}}}
m(\alpha_i^{_{J}})m(\alpha_j^{_{J}})$. 

Consider now a $u$-periodic gauge transformation of the form 
\begin{equation}
m=\exp\left[{2\pi iu}/{P}\cdot\mathrm{diag}(k_1^{_{J}},...,k_1^{_{J}},...,
k_{N_J}^{_{J}},...,k_{N_J}^{_{J}})\right],\label{defm2}
\end{equation}
where the multiplicity of the eigenvalues of
$\mathrm{diag}(k_1^{_{J}},...,k_1^{_{J}},...,k_{N_J}^{_{J}},...,k_{N_J}^{_{J}})$ is chosen to be the 
same as that of $a_\circ^{_{J}}$ so that they commute. According to Remark \ref{rem-amb} by making 
use of such a gauge transformation the eigenvalues $\alpha_i^{_{J}}$ of $a_\circ^{_{J}}$ can 
be shifted by adding the values $2\pi ik_i^{_{J}}/P$ where $k_{i}^{_{J}}\in\mathbb{Z}$ denotes the 
eigenvalue of the ${i}^{th}$ diagonal block of $\mathrm{diag}(k_1^{_{J}},...,k_1^{_{J}}...
k_{N_J}^{_{J}},...,k_{N_J}^{_{J}})$. This, in virtue of (\ref{eig}), yields a discrete shifting 
of $\mathfrak{c}_{ij}$ by adding a term of the form $2\pi ic_2(k_i^{_{J}}-k_{j}^{_{J}})/P$ to the 
eigenvalues $\mathfrak{c}_{ij}$.

Assume now that $\mathfrak{c}_{i'j'}=2\pi ik_{i'j'}/P$ for a fixed set of the values of $i',j'$ with 
$\alpha_{i'}^{_{J}}\not=\alpha_{j'}^{_{J}}$ where $k_{i'j'}\in\mathbb{Z}\setminus\{0\}$.
Then, by choosing the integers $k_{i'}^{_{J}}$ in (\ref{defm2}) so that 
\begin{equation}
k_{i'}^{_{J}}-k_{j'}^{_{J}}=c_2^{-1}k_{i'j'}\label{link}
\end{equation}
all of the eigenvalues $\mathfrak{c}_{i'j'}$ can simultaneously be shifted to be zero. To see that the 
inhomogeneous linear system (\ref{link}) does really have a solution of the needed type recall that
the integers $k_{i'j'}$ are not independent since $c_2[\Im(\alpha_{i'}^{_{J}})-\Im(\alpha_{j'}^{_{J}})]
=2\pi k_{i'j'}/P$, where $\Im(\alpha)$ denotes the imaginary part of $\alpha$. Thereby, the 
coefficient and augmented matrices of (\ref{link}) are always of the same rank. 

It follows then that (\ref{defm}) with $\widetilde m=m_{_{J}}$ and with 
a string $(k_1,...,k_N)$ built up from suitably chosen integers 
$k_{i'}^{_{J}}$, yields a preferred gauge representation $\widetilde A_a$ 
with $\widetilde {\mathfrak{c}}_{i'j'}=0$. \hfill\fbox{}

\begin{remark}\label{r-tow}
Consider now the preferred representation $\widetilde A_a$ yielded by the above described process with 
$\widetilde{\mathfrak{c}}_{i'j'}=0$. By making use of a 
suitable gauge transformation of the form (\ref{defm}) it can be mapped to a non-preferred 
representation so that the eigenvalues $\mathfrak{c}'_{i'j'}=2\pi i k_{i'j'}/P$ are freely specifiable. 
Correspondingly, centered on any preferred gauge representation of the type of $\widetilde A_a$ an 
infinite crystal 
of non-preferred representations can be built up. On the other hand, those preferred representations 
for which neither of the eigenvalues $\mathfrak{c}_{ij}$ with $\alpha_{i}^{_{J}}\not=\alpha_{j}^{_{J}}$
vanishes are always shifted to another preferred representation. Consequently, there can exist YM 
fields such that all of their adapted gauge representations are preferred.
\end{remark} 

\begin{example}
Consider the particular case of $SU(2)$ gauge group. Since $\mathfrak{g}=su(2)$ consists of the
traceless skew-hermitian $2\time2$-matrices $a_\circ^{_{J}}$ must be of the form $a_\circ^{_{J}}
=i\alpha_\circ\mathrm{diag}(1,-1)$ for some $\alpha_\circ\in\mathbb{R}$. Hence, for 
$\kappa_\circ\not=0$, an adapted gauge representation $A_a$ with $A_u^\circ=a_\circ$ is non-preferred 
whenever $\alpha_\circ=\pi k/P$ for some $k\in\mathbb{Z}\setminus\{0\}$. Correspondingly, in this case 
there is a single one-dimensional infinite crystal of non-preferred representations centered on 
the preferred representation $a_\circ\equiv 0$.
\end{example}

\medskip 

Turning back to our general argument assume that $A_a$ is a preferred
gauge representation, with $A_u^\circ=a_\circ$, of the considered 
EYM system. Since the gauge 
transformation $\widetilde{m}$ applied in the proof of the above lemma 
had no $r$-dependence the  gauge source function  
\begin{equation} 
{\mathcal A}=\nabla^aA_{a},\label{ymgauge22} 
\end{equation} 
also vanishes on $\mathcal{O}_i$. 
Moreover, by (\ref{ymgauge1}) 
\begin{equation} 
\left(\frac{\partial A_u}{\partial u}\right)^\circ=0.\label{yau00} 
\end{equation} 
  
From this point one can proceed as follows: Since $g_{AB}$ is negative definite and $(\ /\ )$
is positive definite in virtue of (\ref{ymt})
\begin{equation} 
T_{uu}^\circ=-{1\over 4\pi}\left[(F_{uA}/F_{uB})g^{AB}\right]^\circ\geq0 
\end{equation} 
holds which, according to the argument of Prop. \ref{prop-3.1}, implies that 
\begin{equation} 
F_{uB}^\circ=\left({\partial A_B\over \partial u} 
-{\partial A_u\over\partial x^B}+[A_u,A_B]\right)^\circ=0 \ \ \ {\rm and} \ \ 
\ \left({\partial g_{AB}\over \partial u}\right)^\circ=0.\label{yfub} 
\end{equation} 
Then, in virtue of (\ref{yau00}) and (\ref{yfub}) 
\begin{equation} 
\frac{\partial}{\partial u}\left(\frac{\partial A_B }{\partial u}\right)^\circ+ 
\left[a_\circ,\left(\frac{\partial A_B }{\partial u}\right)^\circ\right]=0. \label{yfuB1} 
\end{equation} 
Any $u$-periodic solution $A_B$ of (\ref{yfuB1}) has also to satisfy 
\begin{equation} 
\left(\frac{\partial A_B }{\partial u}\right)^\circ=0.
\label{yfuB} 
\end{equation} 
Note that (\ref{yfub}) and (\ref{ymt}) imply that 
$T_{uA}^\circ=0$ which by the argument of Prop. \ref{prop-4.1} yields that 
\begin{equation} 
\left( {\partial f\over \partial u}\right) 
^\circ=\left({\partial h_{A}\over \partial u}\right) 
^\circ=0. \label{ybeta2} 
\end{equation} 
It can now be shown that $A_r$ is $u$-independent on $\widetilde{\mathcal{N}}_i$. To see 
this, consider the `$u$'-component of 
(\ref{yme}) in Gaussian null coordinates which on $\widetilde{\mathcal{N}}_i$ reads as 
\begin{equation} 
\frac{\partial F_{ru} }{\partial u}+[a_\circ,F_{ru}]=0.\label{yfur0} 
\end{equation} 
In the case of the considered gauge representation any $u$-periodic solution of (\ref{yfur0}) is also 
$u$-independent on $\widetilde{\mathcal{N}}_i$ and it has to commute with $a_\circ$. 
Hence, in particular, 
\begin{equation} 
\left(\frac{\partial F_{ru} }{\partial u}\right)^\circ= 
\frac{\partial }{\partial u}\left[\frac{\partial A_r }{\partial u}- 
\frac{\partial A_u }{\partial r}\right]+\left[a_\circ,\left(\frac{\partial A_r} 
{\partial u}\right)^\circ\right]=0.\label{yru} 
\end{equation} 
In addition, by differentiating (\ref{ymgauge22}) with respect to $u$ and 
setting $r=0$, we get 
\begin{equation} 
\frac{\partial }{\partial u}\left(\frac{\partial A_r }{\partial u}+ 
\frac{\partial A_u }{\partial r}\right) 
+2\kappa_\circ\left(\frac{\partial A_r }{\partial u}\right)=0.\label{yi} 
\end{equation} 
Then by (\ref{yru}) and (\ref{yi}) we have that on $\widetilde{\mathcal{N}}_i$ 
\begin{equation} 
2\frac{\partial }{\partial u}\left(\frac{\partial A_r }{\partial u}\right) 
+2\kappa_\circ\frac{\partial A_r }{\partial u}+ 
\left[a_\circ,\left(\frac{\partial A_r }{\partial u}\right)\right]=0\label{y0} 
\end{equation} 
which for the case of the selected type of gauge representation has the only periodic solution 
\begin{equation} 
\left(\frac{\partial A_r }{\partial u}\right)^\circ=0. 
\end{equation} 
  
\smallskip 
  
From this point the argument of the proof of Prop. 5.2 can be repeated for the 
case of a YM field with obvious notational changes and the additional 
analysis related to the presence of the second term on the l.h.s. of (\ref{yme}) and the 
third term on the r.h.s. (\ref{ymf}). As we have seen -- compare eqs. (\ref{yfuB1}), 
(\ref{yfur0}) and (\ref{y0}) to the corresponding equations applied to prove Prop. 5.2. -- 
these new terms contribute only a single commutator of $a_\circ$ and the `unknown' of 
the relevant equations and `$\{${\it terms independent of u}$\}$'. 
By an inductive argument, it can also be shown that 
\begin{equation} 
\left(\frac{\partial^n }{\partial r^n}\left[A_u,A_r\right]\right)^\circ= 
\left[a_\circ,\left(\frac{\partial^n A_r}{\partial r^n}\right)^\circ\right]+ 
\{terms\; independent\; of\; u\}.\label{yau01} 
\end{equation} 
Similarly, by induction we get that for the `$u$' 
and `$B$'-components of the second term on the l.h.s. of (\ref{yme}) 
\begin{equation} 
\left(\frac{\partial^n }{\partial r^n}\left[A^a,F_{au}\right]\right)^\circ= 
\left[a_\circ,\left(\frac{\partial^n F_{ru}}{\partial r^n}\right)^\circ\right]+ 
\{terms\; independent\; of\; u\}\label{yau02} 
\end{equation} 
and 
\begin{equation} 
\left(\frac{\partial^n }{\partial r^n}\left[A^a,F_{aB}\right]\right)^\circ= 
\left[a_\circ,\left(\frac{\partial^{n+1} A_{B}}{\partial r^{n+1}}\right)^\circ 
\right]+\{terms\; independent\; of\; u\}\label{yau03} 
\end{equation} 
hold on $\widetilde{\mathcal{N}}_i$. Based on these observations the following can be proven: 
  
\begin{proposition}\label{prop-diffymf} 
Let $(\mathcal{O}_i,g_{ab}\vert_{\mathcal{O}_i})$ be an elementary spacetime region associated 
with an E[M]YM system. Then there exists a gauge potential $A_a:\mathcal{O}_i\rightarrow\mathfrak{g}$ 
so that the $r$-derivatives of the functions $f,h_A$ and $g_{AB}$ and also of the components 
$A_u,A_r,A_B$ up to any order are $u$-independent on $\widetilde{\mathcal{N}}_i$.
\end{proposition} 
  
\begin{remark}\label{r-disk} 
It is also important to know what kind of symmetries are adopted by an
EYM system associated with a non-preferred $u$-periodic 
gauge representation. Let $A_a$ be such a gauge representation so that $A_u^\circ=a_\circ$ is
$u$-independent. Since $A_a$ is non-preferred there must exist $k\in \mathbb{Z}\setminus\{0\}$ 
so that $2\pi i k/P$ is an eigenvalue of $\mathfrak{C}(a_\circ;\kappa_\circ,c_1,c_2)$.  
Recall now that the equations (\ref{yfuB1}), 
(\ref{yfur0}) and (\ref{y0}) possess, instead of the above $u$-independent solutions, 
$u$-periodic solutions with smallest positive period $P/\vert k\vert$. These solutions in 
the succeeding equations always yield periodic `forcing terms'. Therefore these equations 
-- and also all the succeeding ones at higher levels of the corresponding hierarchy --  possess  
$u$-periodic solutions, with smallest positive period $P/\vert k\vert$. (The last claim follows 
from Theorem 2.3.6 of \cite{farkas} since the all the relevant functions are ensured to 
be bounded by their original $u$-periodicity.) Consequently, in case of a non-preferred gauge 
representation only the $u$-periodicity of the restrictions of the $r$-derivatives of the functions 
$f,h_A$ and $g_{AB}$ and also that of the components $A_u,A_r,A_B$ onto 
$\widetilde{\mathcal{N}}_i$ can be demonstrated. Nevertheless, we would like to emphasize that 
by making use of the `discrete shifting freedom' the value of $k$ can be adjusted to be an arbitrary 
integer. Hence, by ranging through all the associated non-preferred representations at least the 
$u$-independentness of the metric functions could be shown even though we had no preferred gauge
representation. 
\end{remark}

\begin{remark}\label{r-four} 
In the case of asymptotically flat static spherically symmetric $SU(2)$ YM black 
holes a gauge potential with vanishing $a_\circ$ can always be chosen. However, numerical 
studies indicated that even in the static case the YM black hole spacetimes need 
not to be spherically symmetric \cite{kk1,kk2}. Moreover,  perturbative analyses  
showed that they need not even to be axially symmetric \cite{rw,vs,bh}. In virtue of Remarks
\ref{r-tow} and \ref{r-disk} it would be important to make it clear what an extent the relevant 
gauge freedom had been exhausted in arriving to these conclusions. Clearly, by an adaptation 
of the framework of the present paper analytic studies of the corresponding YM configurations 
could also be carried out at least in a sufficiently small neighborhood of the horizon. 
\end{remark} 

\begin{remark} 
By a straightforward combination of the arguments that apply for the 
separated cases of a scalar field, a Higgs field and a Yang-Mills field 
it can be shown that Prop. \ref{prop-diffmf} and Prop. \ref{prop-diffymf} 
generalize to E[M]YMd and EYMH systems. To see this consider 
an E[M]YMd, resp. an EYMH, system given by the Lagrangian 
$\mathcal{L}=R+\mathcal{L}_{_{\mathrm{matter}}}$ with
\begin{equation} 
\mathcal{L}_{_{\mathrm{matter}}}^{^{_{[M]YMd}}}=
-e^{2\gamma_{_{d}}\psi}(F_{ef}/F^{ef})+2\nabla^e\psi\nabla_e\psi, 
\end{equation} 
resp.
\begin{equation} 
\mathcal{L}_{_{\mathrm{matter}}}^{^{_{YMH}}}=-(F_{ef}/F^{ef})+
2[(\mathcal{D}^e\psi/\mathcal{D}_e\psi)-V(\psi)], 
\end{equation} 
where $R$ is the 4-dimensional 
Ricci scalar, $\psi$ stands for the (real) dilaton field and 
$\gamma_{_{d}}$ is the dilaton coupling constant\footnote{% 
This Lagrangian with $\gamma_{_{d}}=1$ reproduces the usual `low energy' 
Lagrangian obtained from string theory and it reduces to the pure E[M]YM 
Lagrangian with $\gamma_{_{d}}=0$ and $\psi\equiv const$.}, resp. for
the Higgs field with $\mathcal{D}_a\psi=\nabla_a \psi-[A_a,\psi]$ and
with a sufficiently regular but otherwise arbitrary gauge invariant
potential $V(\psi)$, $F_{ab}$ is the YM field strength. 
The validity of our statement is based on the observations that 
the energy-momentum tensor for such an E[M]YMd, resp. EYMH,  system reads as 
\begin{equation} 
T_{ab}^{^{_{[M]YMd}}}=-\frac{1}{4\pi}\left[e^{2\gamma_{_{d}}\psi}
(F_{ae}/{F_b}^{e})-\nabla_a\psi\nabla_b\psi+\frac{1}{4}g_{ab} 
\mathcal{L}_{_{\mathrm{matter}}}^{^{_{[M]YMd}}}\right], 
\end{equation} 
resp.
\begin{equation} 
T_{ab}^{^{_{YMH}}}=-\frac{1}{4\pi}\left[(F_{ae}/{F_b}^{e})
-(\mathcal{D}_a\psi/\mathcal{D}_b\psi)
+\frac{1}{4}g_{ab}\mathcal{L}_{_{\mathrm{matter}}}^{^{_{YMH}}}\right],
\end{equation} 
and also that the field equations are 
\begin{equation} 
\nabla^a\nabla_a\psi+\frac{\gamma_{_{d}}}{2}e^{2\gamma_{_{d}}\psi}
(F_{ef}/F^{ef})=0 
\end{equation} 
\begin{equation} 
\nabla^aF_{ab}+\left[A^a,F_{ab}\right]+2\gamma_{_{d}}{F^a}_{b}
\nabla_a\psi=0,\label{ymde} 
\end{equation} 
resp. 
\begin{equation} 
\nabla^a\mathcal{D}_a\psi-[A^a,\mathcal{D}_a\psi]+\frac{1}{2}\frac{\partial
V}{\partial \psi} =0
\end{equation} 
\begin{equation} 
\nabla^aF_{ab}+\left[A^a,F_{ab}\right]-[\psi,\mathcal{D}_b\psi]=0.\label{ymhe} 
\end{equation} 
By making use of these relations, along with a suitable inductive argument, in case 
of a preferred gauge representation, first  
the $u$-independentness of the $r$-derivatives of gauge potential components -- 
and, in turn, that of $F_{ab}$, $T_{ab}$ and of $g_{ab}$ -- can be demonstrated. Then 
the same order of $r$-derivative of dilaton, resp. Higgs, field, $\psi$, can also be proven to 
be constant along the generators of $\widetilde{\mathcal{N}}_i$.

A similar reasoning does apply in the case of a complex Higgs field
$\psi$ in the fundamental representation of $G$ where the Higgs part
of the Lagrangian is given as
\begin{equation} 
\mathcal{L}_{_{\mathrm{matter}}}^{^{_{Higgs}}}=
2[(\mathcal{D}^e\psi)^*(\mathcal{D}_e\psi)-V(\psi^*\psi)], 
\end{equation} 
with $\mathcal{D}_a\psi=\nabla_a \psi-iA_a\psi$. 
\end{remark}

\begin{corollary}
Let  $(\mathcal{O}_i,g_{ab}\vert_{\mathcal{O}_i})$ be an elementary 
spacetime region associated with an E[M]YMd or an EYMH system as they
were specified above. Then there exists a gauge potential 
$A_a:\mathcal{O}_i\rightarrow\mathfrak{g}$ so that  the $r$-derivatives of the functions
$f,h_{A},g_{AB}$ and $\psi$ and also that of the components $A_u,A_r,A_B$ up to any order are 
$u$-independent on $\widetilde{\mathcal{N}}_i$. 
\end{corollary} 

In virtue of Props. 5.1 - 5.3,  Cors. 5.1 - 5.2 and of the above
remark we have that in the case of an analytic EKG, E[nA]H, E[M]YMd 
or EYMH system $(\partial/\partial u)^a$ is a Killing vector field
in a neighborhood of $\widetilde{\mathcal{N}}_i$. In addition, by making 
use of the argument applied in Remark 3.2 of \cite{frw} the local Killing fields induced by 
the maps $\psi_i:\mathcal{O}_i\rightarrow \mathcal{U}_i$ can be shown to patch together to 
a global Killing field on a neighborhood of $\mathcal{N}$. Thereby we have the following:

\begin{corollary} 
Let $(M,g_{ab})$ be an analytic EKG, E[nA]H, E[M]YMd or EYMH spacetime 
of class A or B. Then there 
exists a Killing vector field $k^a$ in an open neighborhood, $\mathcal{V}$ of $\mathcal{N}$ 
so that it is normal to $\mathcal{N}$ and the matter fields are also invariant in $\mathcal{V}$. 
\end{corollary}
  
\section{Existence of a horizon Killing vector field}\label{sec-ext} 
\setcounter{equation}{0} 
  
This section is to show that  Prop. 4.1 and Theor. 4.1 of \cite{frw} 
generalize from Einstein-Maxwell spacetimes to EKG, E[nA]H and 
(in parts) also to E[M]YMd  and EYMH configurations. 
The matter fields of these coupled Einstein-matter systems will simply
be denoted by $(0,l_j)$ type tensor fields, $\mathcal{T}_{_{(j)}}$. 
Accordingly, $\mathcal{T}_{_{(1)}}$ stands for a single field $\psi$ 
in case of EKG and E[nA]H systems whereas $\mathcal{T}_{_{(1)}}$ and  
$\mathcal{T}_{_{(2)}}$ denote the dilaton or Higgs field $\psi$ 
and the vector potential $A_a$, respectively, in case of 
E[M]YMd or EYMH systems. 

\begin{proposition}\label{theor-ext}
Let $(\mathcal{O}_i,g_{ab}\mid _{\mathcal{O}_i})$ be an elementary spacetime region 
associated with either an EKG, E[nA]H, E[M]YMd or an EYMH spacetime of class B such that 
$\kappa_{\circ}>0$. Then the followings hold: 

(i) There exists an open neighborhood, $\mathcal{O}_i''$, of $\widetilde{\mathcal{N}}_i$ 
in $\mathcal{O}_i$ such that $(\mathcal{O}_i'',g_{ab}\mid _{\mathcal{O}_i''})$ can be 
extended to a smooth spacetime, $(\mathcal{O}^*,g^*_{ab})$, that possesses a bifurcate null
surface, $\widetilde{\mathcal{N}}^*$---i.e., $\widetilde{\mathcal{N}}^*$ is the union of 
two null hypersurfaces, $\mathcal{N}^*_1$ and $\mathcal{N}^*_2$, which intersect on a
2-dimensional spacelike surface, $S$---such that $\widetilde{\mathcal{N}}_i$ corresponds 
to the portion of $\mathcal{N}^*_1$ that lies to the future of $S$ and $I^+[S] =
\mathcal{O}_i'' \cap I^+[\widetilde{\mathcal{N}}_i]$.

(ii) The fields $\mathcal{T}_{_{(j)}}$ also extend smoothly to tensor fields 
$\mathcal{T}_{_{(j)}}^*$ on $\mathcal{O}^*$ in the case of EKG, E[nA]H and EMd spacetimes. 
In general, preferred gauge potentials of E[M]YMd or EYMH spacetimes blow up at $\mathcal{N}^*_2$ 
so they can be smoothly extended  only onto $\mathcal{O}^*\setminus\mathcal{N}^*_2$. 

(iii) $k^a=(\partial/\partial u)^a$  extends smoothly from $\mathcal{O}_i''$ to a vector 
field $k^a{}^*$ on $\mathcal{O}^*$. In addition, $\mathcal{L}_{k^*}\mathcal{T}_{_{(j)}}^*$
can be defined everywhere in $\mathcal{O}^*$ for any of the considered systems, moreover, $\mathcal{L}_{k^*}g_{ab}^*$ and 
$\mathcal{L}_{k^*}\mathcal{T}_{_{(j)}}^*$ vanish on $\widetilde{\mathcal{N}}^*$.
\end{proposition}

\noindent{\bf Proof}{\ } The justification of the smooth extendibility of the spacetime 
geometry $g_{ab}$ is almost identical to that of the first part of Prop. 4.1 with the 
following distinction: To demonstrate that in $\mathcal{O}_i$, (which has now the same
properties as $\mathcal{O}_i'$ had in \cite{frw}) the spacetime metric $g_{ab}$ can be 
decomposed as
\begin{equation}
g_{ab} = g^{(0)}_{ab} + \gamma_{ab}
\end{equation}
where, in the Gaussian null coordinates of Prop. 4.1, the
components, $g^{(0)}_{\mu \nu}$, of $g^{(0)}_{ab}$ are independent of
$u$, whereas the components, $\gamma_{\mu \nu}$, of $\gamma_{ab}$ and
all of their derivatives with respect to $r$ vanish at $r=0$ we need to
refer to Props. 5.1 - 5.3 and Cors. 5.1 - 5.2 instead of eq. (3.2) of \cite{frw}.

\smallskip

In turning to the proofs of the statements of $(ii)$ and $(iii)$ note first that in 
virtue of Props. 5.1 - 5.3 and Cors. 5.1 - 5.2 the fields $\mathcal{T}_{_{(j)}}$ can be 
decomposed as 
\begin{equation}
\mathcal{T}_{_{(j)}} = \mathcal{T}_{_{(j)}}^{(0)}+ \widehat{\mathcal{T}}_{_{(j)}},
\end{equation}
where the components of $\mathcal{T}_{_{(j)}}^{(0)}$, in Gaussian null coordinates of 
Prop. 4.1, are independent of $u$, while the components  of $\widehat{\mathcal{T}}_{_{(j)}}$
and all of their $r$-derivatives vanish at $r=0$. Then an argument, analogous to that 
applied in \cite{frw} to show the smooth extendibility of $\gamma_{ab}$, can be used to
demonstrate that $\widehat{\mathcal{T}}_{_{(j)}}$ extend smoothly to 
$\widehat{\mathcal{T}}_{_{(j)}}^*$ on $\mathcal{O}^*$ so that the components of 
$\widehat{\mathcal{T}}_{_{(j)}}^*$ and all of their Kruskal coordinate derivatives are zero
on $\widetilde{\mathcal{N}}^*$. In particular, this extension can be done so that 
$\widehat{\mathcal{T}}_{_{(j)}}^*$ are invariant under the action of the `wedge reflection'
isometry defined by $(U,V)\rightarrow (-U,-V)$ on $(\mathcal{O}^*,g^*_{ab})$. 

Thereby the fields $\mathcal{T}_{_{(j)}}$ themselves extend smoothly to $\mathcal{O}^*$ 
whenever the fields $\mathcal{T}_{_{(j)}}^{(0)}$ do. It is straightforward to show that a 
$u$-independent scalar field (or a set of scalar fields), associated with a Klein-Gordon, 
Higgs or dilaton field, represented by $\psi^{(0)}$ extends smoothly to a field 
$\psi^{(0)}{}^*$ on $\mathcal{O}^*$. Hence $\psi$ smoothly extends to 
\begin{equation}
\psi^*=\psi^{(0)}{}^*+\widehat{\psi}^*
\end{equation}
that is constant along the generators of $\widetilde{\mathcal{N}}^*$. This, along with the 
fact that $k^a=(\partial/\partial u)^a$ extends smoothly to 
\begin{equation}
k^a{}^*=\kappa_\circ\left[ U\left(\frac{\partial}{\partial U}\right)^a-
V\left(\frac{\partial}{\partial V}\right)^a\right],\label{deftk}
\end{equation}
implies that $\mathcal{L}_{k^*}\psi^*$ vanishes throughout $\widetilde{\mathcal{N}}^*$.

Consider now the extendibility of the $k^a{}^*$-invariant part $A_a^{(0)}$ of a preferred gauge
representation of a Maxwell-Yang-Mills field. In Gaussian null coordinates $A_a^{(0)}$ can be 
given as 
\begin{equation} 
A_a^{(0)}=A_a^{(0,0)}+r\cdot A_a^{(0,1)}+ O_a(r^2).
\end{equation} 
Taking account of the transformation between the Gaussian null coordinates and the 
generalized Kruskal coordinates (see eqs. (24) and (25) of \cite{rw1}) we get
that there exist smooth functions $A_{\alpha}^{^{(0,I)}}$ with $I=0,1$ so that 
\begin{equation} 
\left. 
\begin{array}{rl} 
{A}_{U}^{(0)}&=(\kappa_\circ U)^{-1}a_\circ +V\left\{A_{U}^{^{(0,0)}} 
(x^3,x^4)+UVA_{U}^{^{(0,1)}}(UV,x^3,x^4)\right\}\\ 
{A}_{V}^{(0)}&=\phantom{(\kappa_\circ U)^{-1}a_\circ +}\ 
U\left\{A_{V}^{^{(0,0)}}(x^3,x^4)+UVA_{V}^{^{(0,1)}}(UV,x^3,x^4)\right\}\\ 
{A}_{x^B}^{(0)}&=\phantom{(\kappa_\circ U)^{-1}a_\circ +U\{\ }
A_{x^B}^{^{(0,0)}}(x^3,x^4)+UVA_{x^B}^{^{(0,1)}}(UV,x^3,x^4).\\ 
\end{array} 
\right.  \label{NvF} 
\end{equation}   
Clearly $A_a^{(0)}$ extends smoothly to a wedge reflection invariant field 
$A_a^{(0)}{}^*$ on $\mathcal{O}^*$ in the case of a Maxwell or a Yang-Mills field 
with vanishing $a_\circ$. Whenever $a_\circ\not\equiv 0$ an extension of  
$A_a^{(0)}{}^*$ of $A_a^{(0)}$ can be defined only on $\mathcal{O}^*\setminus\mathcal{N}_2^*$ 
because then the first term of the r.h.s. of the first equation of (\ref{NvF}) blows up at 
$U=0$, i.e. at $\mathcal{N}_2^*$. 

On contrary to the `parallelly propagated' blowing up of the gauge potential $A_a^{(0)}$ at  $U=0$ 
its Lie derivative, $\mathcal{L}_{k}A_a^{(0)}$, with respect to $k^a$ is regular on $\mathcal{O}_i$ 
and vanishes on $\widetilde{\mathcal{N}}_i$. To see this recall that 
$\mathcal{L}_{k}{A}_{a}^{(0)}=k^e\partial_e A_{a}^{(0)}+A_{e}^{(0)}\partial_ak^e.$ 
Moreover, by (\ref{deftk}) $\partial_\alpha k^\beta=\kappa_ 
\circ({\delta^\beta}_U{\delta^U}_\alpha-{\delta^\beta}_V{\delta^V}_\alpha)$ 
in $\mathcal{O}_i$, which implies that 
\begin{equation} 
\mathcal{L}_{k}A_{\alpha}^{(0)}=\kappa_\circ\left\{U{\partial A_\alpha^{(0)}\over\partial U} 
-V{\partial A_\alpha^{(0)}\over \partial V}+ \left(A_U^{(0)}{\delta^U}_\alpha- 
A_V^{(0)}{\delta^V}_\alpha\right)\right\} \label{lieA} 
\end{equation} 
holds. Since the singular terms in (\ref{lieA}) compensate each other $\mathcal{L}_{k}A_a^{(0)}$ 
is well-defined in $\mathcal{O}_i$ and vanishes on $\widetilde{\mathcal{N}}_i$. In turn, 
$\mathcal{L}_{k^*}A_a^*$ can be defined everywhere in $\mathcal{O}^*$ and it follows from 
(\ref{lieA}) that $\mathcal{L}_{k^*}A_a^*$ vanishes whenever either 
$U=0$ or $V=0$, i.e. on  $\widetilde{\mathcal{N}}^*$. 

The wedge reflection symmetry of the fields $\mathcal{T}_{_{(j)}}^*$ ensure that the coupled
Einstein-matter fields equations are satisfied everywhere in the common domain of their 
definition. \hfill\fbox{} 

\begin{remark} 
Since $\mathcal{L}_{k^*}g_{ab}^*=0$ on $\widetilde{\mathcal{N}}^*$ the bifurcate 
null surface $\widetilde{\mathcal{N}}^*$ is expansion and shear free. In addition, 
by an argument analogous to the one applied in Remark \ref{rem-gs}, it can be shown that the vector 
field $k^a{}^*$ is a repeated principal null vector field of the Riemann tensor on 
$\widetilde{\mathcal{N}}^*_2$, i.e. we have that $\Psi_0=\Psi_1=\Phi_{00}=\Phi_{01}=0$ there. 
\end{remark} 
  
\begin{remark}\label{rem-gs2} 
It is tempting to conclude that $\widetilde{\mathcal{N}}^*$ is completely regular, at least 
in case of the considered systems. On contrary to this, some components of the curvature tensor 
can blow up along the generators of  $\widetilde{\mathcal{N}}^*$  in certain situations.
To see this recall that by the relation of the generalized Kruskal coordinates, introduced 
in \cite{rw1,rw2}, and the Gaussian null coordinates we have that on $\widetilde{\mathcal{N}}_i$ 
\begin{equation} \left. 
\begin{array}{rl} 
R_{UVAV}&={ \kappa_\circ}^{-1}U\cdot R_{urAr}\label{urAr}\\ 
R_{AVBV}&=U^2\cdot R_{ArBr}.\label{ArBr} 
\end{array} 
\right. 
\end{equation} 
Consequently, whenever 
either of the curvature tensor components $R_{urAr}$ or $R_{ArBr}$ is not identically zero 
along one of the generators of $\widetilde{\mathcal{N}}_i$ then the corresponding component, 
$R_{UVAV}$ or $R_{AVBV}$, blows up while $U$ tends to infinity.\footnote{%
The divergence rate of the components $R_{UVAV}$ and $R_{AVBV}$ is exactly the reciprocal 
of the power law found for the strongest possible blowing up of the tidal force tensor 
components along an incomplete maximal causal geodesics upon approaching to the 
associated spacetime `singularity' (see e.g. \cite{kr} and references therein).} Such a 
`parallelly propagated' curvature blowing up can be shown to happen along the generators of 
the event horizon of the `naked black hole' spacetimes -- including e.g. certain EMd systems -- 
studied in \cite{hr1,hr2}. The curvature blows up at `$U=\infty$', i.e. infinitely far from 
the bifurcation surface, and this can occur even though $\widetilde{\mathcal{N}}_i$ is a 
Killing horizon. 
\end{remark} 
    
The remaining part of this section is devoted to the presentation of our main 
result: 

\begin{theorem} 
Let $(M,g_{ab})$ be a smooth EKG, E[nA]H, E[M]YMd or EYMH spacetime of class B so that the generators 
of ${\mathcal{N}}$ are past incomplete. Then there exists an open neighborhood, $\mathcal{V}$ of 
$\mathcal{N}$ such that in $J^+[\mathcal{N}] \cap \mathcal{V}$ there exists a smooth Killing vector 
field $k^a$ which is normal to $\mathcal{N}$. Furthermore, the matter fields are also invariant, i.e. 
$\mathcal{L}_k \mathcal{T}_{_{(j)}}$ vanish, in $J^+[\mathcal{N}]\cap\mathcal{V}$.
\end{theorem} 
  
\noindent{\bf Proof}{\ } The proof of the above statement is almost identical with that of 
Theor. 4.1 of \cite{frw}. The only distinction is that in showing the existence of a Killing 
vector field in the domain of dependence of 
$\widetilde{\mathcal{N}}^*=\mathcal{N}^*_1\cup\mathcal{N}^*_2$ in the extended spacetime 
$\mathcal{O}^*$ Prop. B.1 of \cite{frw} has to be replaced by the following argument: 

In view of Prop. \ref{theor-ext} for EKG, E[nA]H, E[M]YMd and EYMH systems $\mathcal{L}_{k^*}g_{ab}$ and
$\mathcal{L}_{k^*}\mathcal{T}_{_{(j)}}^*$  vanish on $\widetilde{\mathcal{N}}^*$. By referring to 
Theors. 3.1, 4.1 and Remark 4.1 of \cite{r} it can be shown then that the solution $K^a$ of  
\begin{equation} 
\nabla ^e\nabla _eK^a+R_{~e}^aK^e=0  \label{Laplrk} 
\end{equation} 
with initial data $[K^a]=k^a{}^*\vert_{\widetilde{\mathcal{N}}^*}$ is a Killing vector field 
(at least in a sufficiently small neighborhood of $\widetilde{\mathcal{N}}^*$) in the domain
of dependence of $\widetilde{\mathcal{N}}^*$ so that the matter fields are also invariant there.
The only non-trivial step related to the justification of the above claim is to show the 
uniqueness of solutions to the coupled linear homogeneous wave equations satisfied 
by $\mathcal{L}_{k^*} g_{ab}^*$ and $\mathcal{L}_{k^*} \mathcal{T}_{_{(j)}}^*$ in case of
EYMd systems with a preferred gauge representation possessing a p.p. blowing up at $\mathcal{N}^*_2$. 
The key 
observation here is that, due to the regularity of $\mathcal{L}_{k^*} A^*_a$ in $\mathcal{O}^*$ 
(see Prop. \ref{theor-ext}) and also to the smoothness of $g^*_{ab}$, the principal parts of 
the relevant linear homogeneous wave equations are regular. Thereby the energy estimate 
method of the standard uniqueness argument can also be adapted to the present case. 

It follows then that by restricting to $\mathcal{O}_i$ we obtain a Killing field $K^a$ (with 
$\mathcal{L}_K \mathcal{T}_{_{(j)}}=0$) on a one-sided neighborhood of $\widetilde{\mathcal{N}}_i$
of the form $J^+[\widetilde{\mathcal{N}}_i] \cap\widetilde{\mathcal{V}}_i$, where 
$\widetilde{\mathcal{V}}_i$ is an open neighborhood of $\widetilde{\mathcal{N}}_i$.

From this point the proof is identical to that of Theor. 4.1 of \cite{frw}.\hfill\fbox{} 

\medskip  

In view of Prop. 3.1 of \cite{frw} we also have the following

\begin{corollary}\label{c-c}
Let $(M,g_{ab})$ be a smooth EKG, E[nA]H, E[M]YMd or EYMH spacetime of class A so that the generators 
of the event horizon $\mathcal{N}$ are past incomplete. Then there exists an open neighborhood, 
$\mathcal{V}$ of $\mathcal{N}$ such that in $J^+[\mathcal{N}] \cap \mathcal{V}$ there exists a smooth 
Killing vector field $k^a$ which is normal to $\mathcal{N}$. Furthermore, the matter fields are also 
invariant in $J^+[\mathcal{N}] \cap \mathcal{V}$. 
\end{corollary}
  
\section{Concluding remarks}\label{sec-final} 
\setcounter{equation}{0} 
  
In virtue of Cor. \ref{c-c} any EKG, E[nA]H, E[M]YMd or EYMH black hole spacetime of 
class A has to admit a horizon Killing vector field. In the smooth non-static case, the domain 
on which the existence of this Killing vector field 
is guaranteed is `one-sided' and it is contained by the black hole region. On the other hand, e.g. 
in the black hole uniqueness arguments the existence of this Killing vector field in the 
exterior region is what is relevant. Therefore, further investigations will be needed to show 
that the horizon Killing vector field `extends' to the domain of outer communication side. 

\smallskip 
  
Note that in the case of spacetimes  of class B with a compact Cauchy horizon it is 
completely satisfactory to show the existence of a Killing vector field 
on the Cauchy development side, $J^+[\mathcal{N}] \cap \mathcal{V}$.
According to our results the presence of a compact Cauchy horizon 
ruled by closed null geodesics is simply an artifact of a spacetime 
symmetry. In turn, the 
presented results support the validity of the strong cosmic censor 
hypotheses by demonstrating the non-genericness of spacetimes 
possessing such a compact Cauchy horizon. 
  
\smallskip 
  
It is important to emphasize that to have a complete proof of the strong cosmic 
censorship conjecture for spacetimes with a compact Cauchy 
horizon the case of non-closed generators also has to be investigated. 
  
\smallskip 
  
Remember that our result concerning the extendibility of an elementary spacetime region 
is based on the assumption that the horizon is non-degenerate. Therefore in the 
non-analytic case spacetimes with a degenerate horizon are out of our scope. 
Obviously, to show the existence of a horizon Killing field in case of a 
smooth spacetime with geodesically complete horizon would also deserve 
further attentions. 
  
\smallskip 
  
We would like to emphasize that in sections 3 and 4 no use of the particular form 
of the Einstein's equations and the matter field equations was made. 
Thereby, the results contained by these sections generalize straightforwardly 
to those covariant metric theories of coupled gravity-matter systems within 
which the Einstein tensor $G_{ab}=R_{ab}-{1\over 2}Rg_{ab}$ satisfies the following 
generalized form of the dominant energy condition: For all future directed timelike 
vector $\xi^a$ the contraction ${G^a}_b\xi^b$ is a future directed timelike or null vector.  

\smallskip 

In particular, the `zeroth law' of black hole thermodynamics can be 
shown to be valid (see Remark \ref{rem-zero}) for arbitrary black hole spacetimes of class A. 
It is important to emphasize that the relevant argument rests only on the use of the above
generalized form of the dominant energy condition and the event horizon is not assumed to be 
a Killing horizon  as it is usually done in either of the standard arguments. 

\smallskip 
  
It is also of obvious interest to know what sort of event horizon can be associated, in this 
general setting, with a black hole spacetime of class A. In general, there seems to
be no way to show the $u$-independence of the $r$-derivatives of the metric functions
up to arbitrary order. Hence, a full generalization of Prop. 6.1 will probably not be available. 
Nevertheless, the functions $f, h_A$ and $g_{AB}$ have been found to be constant along the 
generators of $\widetilde{\mathcal{N}}_i$. This, whenever $\kappa_\circ\not=0$ along with a 
straightforward adaptation of the argument of Prop. 6.1 (see also Prop. 
4.1 of \cite{frw}), can be used to show that $(\mathcal{O}_i'',g_{ab}\mid _{\mathcal{O}_i''})$ 
extends into a $C^0$ spacetime $(\mathcal{O}^*,{g}_{ab}^*)$ that possesses a bifurcate 
type null hypersurface $\widetilde{\mathcal{N}}^*$. The metric ${g}_{ab}^*$ can be 
ensured to be smooth on $\mathcal{O}^*\setminus \mathcal{N}^*_2$ although it is guaranteed 
only to be continuous through $\widetilde{\mathcal{N}}_2^*$. This argument applies to an 
arbitrary gravity-matter system provided that the generalized form of the  dominant energy 
condition is satisfied. Thereby, this result strengthens the conclusion of \cite{rw1,rw2} 
significantly by demonstrating that the event horizon of a physically reasonable asymptotically 
flat stationary black hole spacetime is either degenerate or it is of bifurcate type. 
    
\section*{Acknowledgments} 
I would like to say thanks to Helmut Friedrich for various helpful suggestions 
and critical comments on a former version of the present work. I also wish 
to thank Masahiro Anazawa, Dmitri V. Gal'tsov and Takeshi Oota for discussions.
This research was supported in parts by the Monbusho Grant-in-aid No. 96369.

\vfill\eject 

\end{document}